\documentclass[sigconf]{acmart}

\usepackage{amsmath}
\usepackage{latexsym}
\usepackage{enumitem}
\usepackage[T1]{fontenc}
\usepackage[utf8]{inputenc}
\usepackage{xspace}
\usepackage{todonotes}
\usepackage{booktabs}
\usepackage{soul}
\usepackage{algorithm}
\usepackage{algorithmic}
\usepackage{newfloat}
\usepackage{listings}
\usepackage{booktabs}
\usepackage{graphicx}
\usepackage{subcaption} 
\usepackage{fontawesome} 
\usepackage{comment}
\usepackage{url}
\usepackage{multirow}
\usepackage{balance}

\definecolor{con}{HTML}{0070C0}
\definecolor{lab}{HTML}{FF0000}
\definecolor{tvt}{HTML}{FF7D18}
\definecolor{snp}{HTML}{FFFF00}
\definecolor{b60}{HTML}{7030A0}
\definecolor{ase}{HTML}{38C177}
\definecolor{lch}{HTML}{AB7942}

\copyrightyear{2022}
\acmYear{2022}
\setcopyright{acmlicensed}
\acmConference[WebSci '22]{14th ACM Web Science Conference 2022}{June 26--29, 2022}{Barcelona, Spain}
\acmBooktitle{14th ACM Web Science Conference 2022 (WebSci '22), June 26--29, 2022, Barcelona, Spain}
\acmPrice{15.00}
\acmDOI{10.1145/3501247.3531543}
\acmISBN{978-1-4503-9191-7/22/06}

\begin{document}

\title[The Spread of Propaganda by Coordinated Communities on Social Media]{The Spread of Propaganda by Coordinated Communities\\on Social Media}


\author{Kristina Hristakieva}
\email{christina.hristakieva@gmail.com}
\affiliation{%
    \institution{Sofia University}
    \country{Bulgaria}
}

\author{Stefano Cresci}
\authornote{Corresponding author.}
\email{s.cresci@iit.cnr.it}
\orcid{0000-0003-0170-2445}
\affiliation{%
    \institution{IIT-CNR}
    \country{Italy}
}

\author{Giovanni Da San Martino, Mauro Conti}
\email{{dasan,conti}@math.unipd.it}
\affiliation{%
    \institution{University of Padova}
    \country{Italy}
}

\author{Preslav Nakov}
\email{pnakov@hbku.edu.qa}
\affiliation{%
    \institution{Qatar Computing Research Institute, HBKU}
    \country{Qatar}
}

\renewcommand{\shortauthors}{Hristakieva et al.}

\begin{abstract}
Large-scale manipulations on social media have two important characteristics: (\textit{i})~they use \textit{propaganda} to influence others, and (\textit{ii})~they adopt \textit{coordinated behavior} to spread propaganda and to amplify its impact. Despite the connection between them, these two characteristics have so far been considered in isolation. Here we aim to bridge this gap. In particular, we analyze the spread of propaganda and its interplay with coordinated behavior on a large Twitter dataset about the 2019 UK general election. We first propose and evaluate several measures for quantifying the use of propaganda on Twitter. Then, we investigate the use of propaganda by different coordinated communities that participated in the online debate. The combined analysis of propaganda and of coordination provides evidence about the harmfulness of coordinated communities that would not be available otherwise. For instance, it allows us to identify a harmful politically-oriented community as well as a harmless community of grassroots activists. 
Finally, we compare our measures of propaganda and of coordination to automation scores  (i.e.,~the use of bots) and Twitter suspensions, revealing interesting trends. From a theoretical viewpoint, we introduce a methodology for analyzing several important dimensions of online behavior that are seldom conjointly considered. From a practical viewpoint, we provide new and nuanced insights into inauthentic and harmful online activities in the run up to the 2019 UK general election.
\end{abstract}

\maketitle

\section{Introduction}
\label{sec:intro}

Social media currently represent one of the main channels for information spread and consumption. They are increasingly used by a constantly growing part of the population to maintain active social relationships, to stay informed about socially relevant issues, and to produce content, thus giving voice to the crowds. At the same time, a large portion of online information is biased, misleading, or outright fake~\cite{wardle2017information,dipietro2021new}. Moreover, harmful content can be purposefully shared by malicious actors, and even by unaware users, with the aim to manipulate online audiences, to sow doubt and discord, and to increase polarization~\cite{starbird2019disinfo,cresci2020decade}. The unprecedented importance of social media for information diffusion, combined with their vulnerability to organized misbehavior, sets the stage for online manipulations that can cause tremendous societal repercussions, as witnessed during the US Capitol Hill assault in January 2021~\cite{thelongfuse2021,ng2021coordinating}, and with the rampaging COVID-19 vaccine misinformation~\cite{ferrara2020misinformation,alam-etal-2021-fighting-covid}.

Despite the differences between the broad array of tactics used to carry out online manipulation, many social media campaigns share two fundamental characteristics: (\textit{i})~they use \textit{propaganda} to influence those targeted by the manipulation~\cite{bolsover2017computational}, and (\textit{ii})~they adopt \textit{coordinated actions} to amplify the spread and the outreach of the manipulation and to increase its impact~\cite{nizzoli2021coordinated,tardelli2022detecting}.
Given their importance for online manipulation, each of these characteristics has received scholarly attention. For example, computational linguists developed AI solutions to automatically detect the use of propaganda techniques~\cite{EMNLP19DaSanMartino,dasanmartino2020survey,ACL2021:propaganda:memes}. Similarly, network science frameworks were proposed for detecting coordinated groups of users~\cite{pacheco2020uncovering} and for measuring the extent of coordination~\cite{nizzoli2021coordinated}. Despite recent progress, the study of computational propaganda and coordinated behavior is still in its early stages.\footnote{\url{https://medium.com/1st-draft/how-to-improve-our-analysis-of-coordinated-inauthentic-behavior-a4ec62ce9bff}} As such, and in spite of the interrelationship between propaganda and coordinated behavior, so far, these two aspects have been investigated in isolation. Nonetheless, their combined analysis is promising under multiple viewpoints:
\begin{itemize}
    \item From the propaganda viewpoint, there already exist methods for detecting the use of rhetorical techniques to influence others~\cite{EMNLP19DaSanMartino}. However, there have been no studies to detect the \textit{intent to harm} behind propaganda campaigns~\cite{dasanmartino2020survey}. Notably, coordination between users implies \textit{a shared intent}. Thus, adding coordination information to the analysis of propaganda can contribute to bridging this gap.
    \item From the coordination viewpoint, there already exist methods for detecting coordinated users in social media~\cite{nizzoli2021coordinated,pacheco2020uncovering,weber2021amplifying}. However, distinguishing between \textit{harmless} (e.g.,~activists, fandoms) and \textit{harmful} (e.g.,~botnets, trolls) coordination is still an open challenge~\cite{vargas2020detection}. Propaganda implies the \textit{aim to mislead and to manipulate}. Thus, adding information about propaganda to the analysis of coordination can help detect harmful behavior.
\end{itemize}
Our aim is to combine techniques for the analysis of propaganda and online coordination to draw nuanced insights into (\textit{i})~the spread of propaganda online, (\textit{ii})~the behavior of coordinated communities, and (\textit{iii})~the interplay between propaganda and coordination.

\textbf{Contributions.} We analyze the -- so far unexplored -- interplay between propaganda and coordination in the context of societal online debates. Towards this goal, we adopt a methodological approach grounded on state-of-the-art techniques for detecting propaganda in texts and for measuring coordinated behavior on social media, and we apply it to study a recent and relevant online debate on Twitter, about the 2019 UK general election. 
In particular, we propose and experiment with several measures for quantifying the spread of propaganda by social media users and communities. We further carry out network analysis of coordinated online communities that participated in the electoral debate. Next, we combine our results on propaganda and coordination by comparing the spread of propaganda to the activity of coordinated communities. We also compare our results to clear signs of inauthenticity and harmfulness -- namely, bot scores provided by Botometer~\cite{sayyadiharikandeh2020detection} and Twitter suspensions. Our analysis provides more nuanced results compared to existing work, and it surfaces interesting patterns in the behavior of online communities that would not be visible otherwise. For instance, it allows to clearly identify and to differentiate communities that exhibit opposite behaviors, such as (\textit{i})~a malicious politically-oriented community, characterized by strongly coordinated users that are involved in spreading propaganda, and (\textit{ii})~a grassroots community of activists protesting for women's rights. Our main contributions can be summarized as follows:
\begin{itemize}
    \item We explore the interplay between propaganda and coordination in online debates, which so far has received little attention.
    \item By cross-checking propaganda and coordination, we get better insights into malicious behavior, thus moving in the direction of identifying and studying coordinated inauthentic behaviors (CIB) as well as propaganda campaigns.
    \item Regarding malicious behavior, we draw insights into the interplay between propaganda/coordination and automation/suspensions.
    \item From a practical standpoint, our analysis reveals interesting and nuanced characteristics of several online communities, which were previously unknown.
\end{itemize}

\textbf{Significance.} Our proposed methodology and results contribute to improving our understanding of coordinated harmful behavior. Moreover, insights such as those obtained thanks to our analysis can support platform administrators at enforcing targeted moderation interventions for curbing online harms~\cite{jhaver2021evaluating,trujillo2022make}.

\section{Related Work}
\label{sec:related}

Propaganda and coordinated behavior pose peculiar challenges that mandate different methods, such as natural language understanding for the former, and network science for the latter. Thus, they have been the focus of largely disjoint efforts by different communities.

\subsection{Coordinated Behavior}
\label{sec:related-coord}

Coordinated behavior, be it authentic or not, was introduced as a concept by Facebook in 2018\footnote{\url{http://about.fb.com/news/2018/12/inside-feed-coordinated-inauthentic-behavior/}} and later widely adopted in studies on online manipulation. Given its recency, the computational analysis of coordinated behavior poses several challenges. Some are conceptual: What exactly is coordinated behavior? How many accounts, or how much coordination, is needed for meaningful coordinated behavior to surface? Currently, there are no agreed-upon answers,\footnote{\url{http://slate.com/technology/2020/07/coordinated-inauthentic-behavior-facebook-twitter.html}} which makes computational analysis problematic. In fact, many solutions still require a great deal of manual intervention~\cite{starbird2019disinfo,mirbabaie2021development}.

In the few recent computational frameworks, coordination was defined as an exceptional similarity between a number of users. \citet{nizzoli2021coordinated} proposed a state-of-the-art pipeline organized in six analytical steps, starting with (\textit{i})~selection of a set of users to investigate, (\textit{ii})~selection of a measure for the similarity between users, (\textit{iii})~construction of a weighted user-similarity network, (\textit{iv})~network filtering, (\textit{v})~coordination-aware community detection, and finally, (\textit{vi})~analysis of the discovered coordinated communities. This approach is the only one so far that has proven capable of producing fine-grained estimates of the extent of coordination in the continuous $[0,1]$ range, rather than a binary $\{0,1\}$ classification of coordinated vs. non-coordinated communities. Examples of methods of the latter type include~\cite{pacheco2020uncovering,pacheco2020unveiling,weber2021amplifying,sharma2021identifying,magelinski2021synchronized}. Similarly to~\cite{nizzoli2021coordinated}, \citet{pacheco2020uncovering} built a weighted user-similarity network. Then, they discarded all edges whose weight is below a threshold, and clustered the remaining network to discover coordinated communities. The drawback of this method, and similar ones~\cite{giglietto2020takes}, is the need to specify arbitrary thresholds to distinguish between coordinated  and non-coordinated behavior, thus providing a binary classification of a non-binary and nuanced phenomenon. Moreover, additional arbitrariness might also arise from the network projection and filtering steps, whose choice can significantly affect coordination results~\cite{coscia2019impact}. Other methods do not embed a notion of coordination, but rather propose to apply community detection to weighted user-similarity networks, thus leaving the task of investigating coordinated communities for subsequent analysis~\cite{weber2020s}.

Notably, in all previous work, coordination was detected or measured independently of harmfulness or authenticity. In fact, coordination does not necessarily imply malicious activities: think of online fandoms, or other grassroots initiatives, which are examples of coordinated \textit{harmless and authentic behavior}. \citet{vargas2020detection} evaluated the capabilities of existing systems to distinguish between harmful and harmless coordination, finding unsatisfactory results and highlighting the difficulty of this task. To this end, our results show that the combined analysis of coordination and propaganda allows to draw insights into the harmfulness and the authenticity of online behavior, thus contributing to bridging this scientific gap.

\subsection{Computational Propaganda}
\label{sec:related-prop}

Work on propaganda detection has focused on analyzing textual documents \cite{BARRONCEDENO20191849,EMNLP19DaSanMartino,rashkin-EtAl:2017:EMNLP2017}.
See \cite{dasanmartino2020survey} for a recent survey on computational propaganda detection.
\citet{rashkin-EtAl:2017:EMNLP2017} developed a corpus with document-level annotations with four classes (\emph{trusted}, \emph{satire}, \emph{hoax}, and \emph{propaganda}), labeled using distant supervision: all articles from a given news outlet were assigned the label of that outlet. The news articles were collected from the English Gigaword corpus, which covers reliable news sources, as well as from seven unreliable news sources including two propagandistic ones. They trained a model using word $n$-grams, and found that it performed well only on articles from sources that the system was trained on, and that the performance degraded quite substantially when evaluated on articles from unseen news sources.
\citet{BARRONCEDENO20191849} developed a corpus with two labels (i.e.,~\emph{propaganda} vs. \emph{non-propaganda}) and further investigated writing style and readability level. Their findings confirmed that using distant supervision, in conjunction with rich representations, might encourage the model to predict the source of the article, rather than to discriminate propaganda from non-propaganda. The studies by \citet{Habernal.et.al.2017.EMNLP,Habernal2018b} also proposed a corpus with 1.3k arguments annotated with five fallacies that directly relate to propaganda techniques, including \textit{ad hominem}, \textit{red herring}, and \textit{irrelevant authority}.

A more fine-grained propaganda analysis was done by \citet{EMNLP19DaSanMartino}, who developed a corpus of news articles annotated with the spans of use of eighteen propaganda techniques. They asked to predict the spans of use of propaganda, as well as the specific technique being used, and they further tackled a sentence-level propaganda detection task. 
They proposed a multi-granular gated deep neural network that captures signals from the sentence-level to improve the performance of the fragment-level classifier and vice versa. Subsequently, an online demo \texttt{Prta} was made publicly available \cite{da2020prta}, and there were several shared tasks \cite{da2020semeval,SemEval2021-6-Dimitrov}.

A limitation of this body of work lies in the lack of methods and tools for uncovering orchestrated propaganda campaigns rather than for detecting individual posts or articles that make use of propaganda. Below, we show that our analysis of coordinated behavior contributes to reaching this goal.

\begin{table}[t]
	\small
	\centering
	\begin{tabular}{lrrr}
		\toprule
		\textbf{hashtag} & \textit{users} & \textit{tweets} \\
		\midrule
		\textsf{\#GE2019}	                & 436,356			& 2,640,966			\\
		\textsf{\#GeneralElection19}	    & 104,616			& 274,095			\\
		\textsf{\#GeneralElection2019}	    & 240,712			& 783,805			\\
		\textsf{\#VoteLabour}	            & 201,774			& 917,936			\\
		\textsf{\#VoteLabour2019}	        & 55,703			& 265,899			\\
		\textsf{\#ForTheMany}	            & 17,859			& 35,621			\\
		\textsf{\#ForTheManyNotTheFew}	    & 22,966			& 40,116			\\
		\textsf{\#ChangeIsComing}	        & 8,170			    & 13,381            \\
		\textsf{\#RealChange}	            & 78,285			& 274254			\\
		\textsf{\#VoteConservative}	        & 52,642			& 238,647			\\
		\textsf{\#VoteConservative2019}	    & 13,513			& 34,195			\\
		\textsf{\#BackBoris}	            & 36,725			& 157,434			\\
		\textsf{\#GetBrexitDone}	        & 46,429			& 168,911			\\
		\midrule
		\textbf{total}	                    & 668,312			& 4,983,499		        \\
		\bottomrule
	\end{tabular}
	\caption{Statistics about the data collected via hashtags.}
	\label{tab:dataset-hashtags}
\end{table}

\section{Data}
\label{sec:dataset}

The starting point for our study is the dataset from~\cite{nizzoli2021coordinated}. It contains 11,264,820 tweets about the 2019 UK general election, published by 1,179,659 distinct users. The tweets were collected between November 12, 2019 and December 12, 2019 (i.e.,~the election day) using the Twitter Streaming API. In particular, the dataset contains all tweets that use at least one of the election-related hashtags shown in Table~\ref{tab:dataset-hashtags}. We can see in the table that the hashtags used for data collection include both partisan hashtags as well as neutral ones. The dataset further contains the tweets shared by the two main parties (labour and conservative) and their leaders, as well as the interactions (i.e.,~retweets and replies) with such tweets, as summarized in Table~\ref{tab:dataset-accounts}. The final dataset for this study is the combination of the data shown in Tables~\ref{tab:dataset-hashtags} and \ref{tab:dataset-accounts}, and quoted retweets (not counted in the tables). The dataset from~\cite{nizzoli2021coordinated} is publicly available for research purposes.\footnote{\url{http://doi.org/10.5281/zenodo.4647893}}

\begin{table}[t]
	\small
	\centering
	\begin{tabular}{lrrrr}
		\toprule
		&&& \multicolumn{2}{c}{\textbf{interactions}}\\
		\cmidrule{4-5}
		\textbf{account} & \textit{tweets} && \textit{retweets} & \textit{replies}\\
		\midrule
		\textsf{@jeremycorbyn}  & 788   && 1,759,823    & 414,158   \\
		\textsf{@UKLabour}      & 1,002 && 325,219      & 79,932    \\
		\textsf{@BorisJohnson}  & 454   && 284,544      & 382,237   \\
		\textsf{@Conservatives} & 1,398 && 151,913      & 169,736   \\
		\midrule
		\textbf{total}          & 3,642	&& 2,521,499    & 1,046,063 \\
		\bottomrule
	\end{tabular}
	\caption{Statistics about the data from Twitter accounts.}
	\label{tab:dataset-accounts}
\end{table}

In this work, we extend the above Twitter dataset by also collecting and analyzing the textual content of all the news articles shared during the online electoral debate. To collect data about articles, we first parse the 11M tweets, looking for URLs pointing to news outlets, blogs, or other news Web sites. Out of the entire Twitter dataset, we found 35,976 distinct articles from 3,974 Web sites, that were shared 329,482 times during the data collection period. Finally, we leverage the \textit{newspaper3k} Python package\footnote{\url{http://newspaper.readthedocs.io/en/latest/}} to collect the textual content, together with some metadata about each shared article. We leverage the textual content of the shared articles and tweets to detect the use of propaganda. We further measure the similarities in the user tweeting behaviors to measure coordination.

\begin{figure}[t]
    \centering
    \includegraphics[width=0.8\columnwidth]{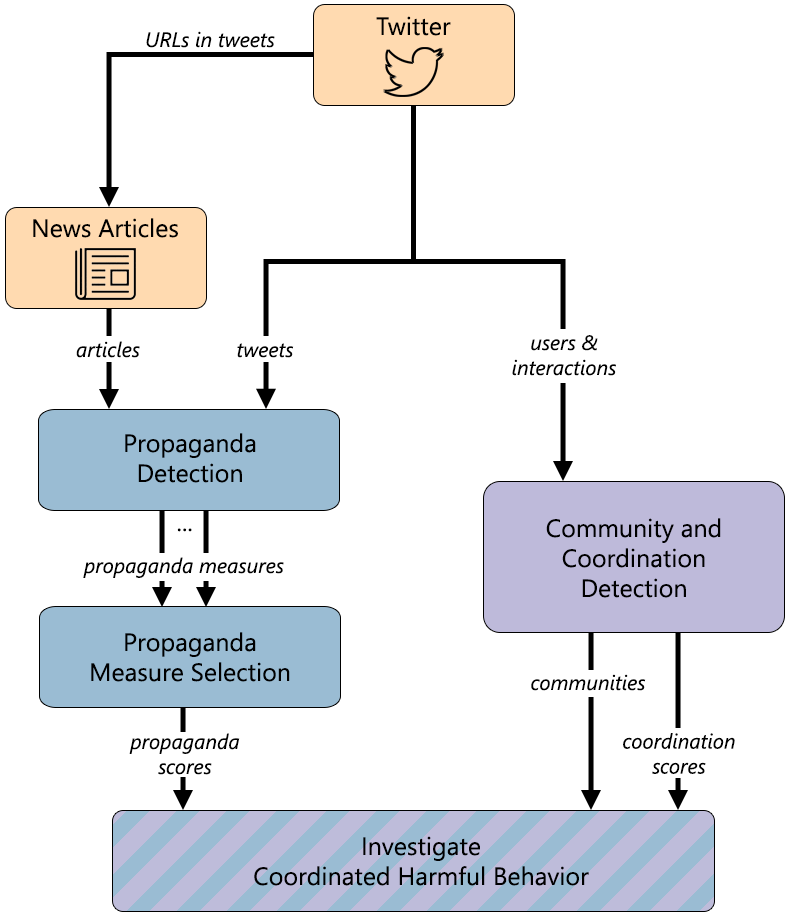}
    \caption{Overview of our approach based on combining the analysis of propaganda and coordination for studying coordinated harmful behavior.}
    \label{fig:method}
\end{figure}

\section{Methods}
\label{sec:method}

Figure~\ref{fig:method} shows our methodological approach to the analysis of coordinated harmful behavior, which has two main building blocks: (\textit{i})~a method for measuring coordination, and (\textit{ii})~a propaganda classifier. As shown in figure, coordination is measured based on user activities and interactions, and results in a coordination score assigned to each user and in the identification of coordinated communities. Instead, propaganda scores are computed from the textual content of tweets and news articles. This process assigns a propaganda score to each user. Finally, each coordinated community is analyzed in terms of the coordination and propaganda scores of its members. The methods for measuring coordination and propaganda are described below.

\subsection{Measuring the Extent of Coordination}
\label{sec:method-coord}

To measure the extent of online coordination, we followed network analysis approaches that have been recently proposed in state-of-the-art studies~\cite{nizzoli2021coordinated,pacheco2020uncovering,weber2021amplifying}, which compute similarities between users and consider exceptional or unexpected similarities as a proxy for coordination. We specifically follow the approach in~\cite{nizzoli2021coordinated}, as it is the only one that produces a coordination score rather then a binary label. 
For the user selection step, we constrained our analysis to \textit{superspreaders}, defined as the top 1\% of the users who shared the most retweets. Despite being only 10,782, superspreaders shared 3.9M tweets, which is 39\% of the tweets and 44\% of the retweets in our dataset. Previous work has shown that focusing on superspreaders is particularly relevant~\cite{pei2014searching}. We measured the similarity between superspreaders in terms of co-retweets, in order to highlight users who frequently reshare the same messages. For each superspreader, we computed a TF.IDF-weighted vector of the tweet IDs that he/she retweeted. Using TF.IDF-weighting discounts viral tweets by influencers and popular users, while emphasizing retweets of unpopular tweets. Then, we computed the similarity between all pairs of superspreaders as the cosine similarity between their corresponding vectors, thus obtaining a weighted undirected user-similarity network. We filtered the network by computing its multiscale backbone, which allows to retain only statistically significant network structures~\cite{serrano2009extracting}. Then, we applied the well-known Louvain community detection algorithm to group users into network communities. Finally, we applied network dismantling, which assigns a coordination score to each user in the network. We carried out the latter step by iteratively removing network edges and nodes based on a moving edge weight threshold. 

\begin{figure}[tbh]
    \centering
    \includegraphics[width=1\columnwidth]{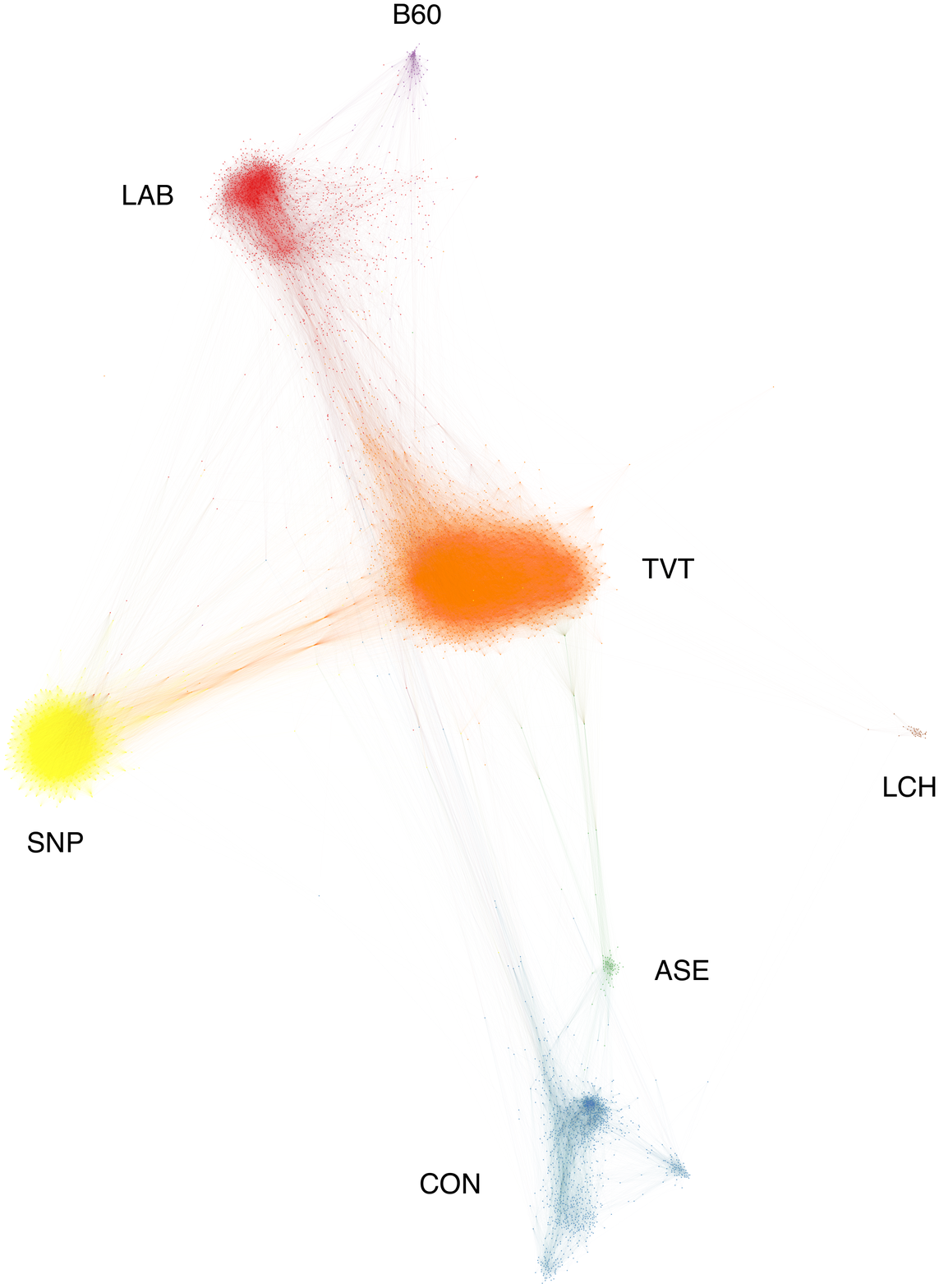}
    \caption{User-similarity network for the 2019 UK electoral debate on Twitter. The different coordinated communities that took part in the online debate are color-coded.}
    \label{fig:network-communities}
\end{figure}

At each iteration, we removed all edges whose raw weight was lower than a threshold, and such that ended up being disconnected from the largest connected component. The threshold increased at each iteration, until the network was completely dismantled, i.e.,~no more connected nodes remained. For each node, we assigned a coordination score as the threshold value that disconnected that node from the rest of the network. We normalized the coordination score in the $[0, 1]$ range, with 1 indicating maximum coordination.

\begin{table*}[t]
    \small
    \centering
    \begin{tabular}{clrcrrrcrrr}
        \toprule
        &&&& \multicolumn{3}{c}{\textbf{articles}} && \multicolumn{3}{c}{\textbf{tweets}} \\
        \cmidrule{5-7} \cmidrule{9-11}
        \multicolumn{2}{l}{\textbf{community}} & \textbf{users} && \multicolumn{1}{c}{\textit{\#}} & \textit{\# distinct} & \textit{(\%)} && \multicolumn{1}{c}{\textit{\#}} & \textit{\# distinct} & \textit{(\%)} \\
        \midrule
        \tikz\draw[black, fill=lab] (0,0) circle (.75ex); & \texttt{LAB} & 5,213 && 79,157 & 5,861 & (7.4\%) && 2,064,041 & 179,601 & (8.7\%) \\
        \tikz\draw[black, fill=con] (0,0) circle (.75ex); & \texttt{CON} & 2,279 && 13,277 & 1,781 & (13.4\%) && 777,537 & 76,191 & (9.8\%) \\
        \tikz\draw[black, fill=tvt] (0,0) circle (.75ex); & \texttt{TVT} & 2,258 && 16,675 & 3,363 & (20.2\%) && 690,900 & 62,058 & (9.0\%) \\
        \tikz\draw[black, fill=snp] (0,0) circle (.75ex); & \texttt{SNP} & 491 && 2,735 & 772 & (28.2\%) && 140,338 & 8,601 & (6.1\%) \\
        \tikz\draw[black, fill=b60] (0,0) circle (.75ex); & \texttt{B60} & 296 && 3,231 & 789 & (24.4\%) && 139,988 & 9,663 & (6.9\%) \\
        \tikz\draw[black, fill=ase] (0,0) circle (.75ex); & \texttt{ASE} & 107 && 706 & 396 & (56.1\%) && 32,887 & 4,723 & (14.4\%) \\
        \tikz\draw[black, fill=lch] (0,0) circle (.75ex); & \texttt{LCH} & 101 && 150 & 57 & (38.0\%) && 28,970 & 2,230 & (7.7\%) \\ [0.75em]
        \multicolumn{2}{l}{overall} & 10,745 && 116,205 & 9,960 & (8.6\%) && 3,886,382 & 343,750 & (8.9\%) \\
        \bottomrule
    \end{tabular}
    \caption{Statistics about the coordinated communities that took part in the 2019 UK electoral debate on Twitter: number of users and shared articles/tweets.\label{tab:communities}}
\end{table*}

\subsection{Propaganda Detection}
\label{sec:method-prop}

In order to assess whether an input piece of text is propagandistic, we used \texttt{Proppy}, a state-of-the-art propaganda detection system that achieved an $F_1$ score of 0.83 on a reference benchmark dataset, outperforming several rival approaches~\cite{BARRONCEDENO20191849}. It uses a maximum entropy binary classifier with L$_2$ regularization to discriminate propagandistic vs. non-propagandistic texts. Proppy was trained on the \texttt{QProp} corpus, which includes 51k news articles from 94 propagandistic and from 10 non-propagandistic news sources.\footnote{From the \textit{Media Bias/Fact Check} website: \url{http://mediabiasfactcheck.com}}

\texttt{Proppy} represents the input text as a set of features, including (\textit{i})~TF.IDF-weighted $n$-grams, (\textit{ii})~frequency of specific words from a number of lexicons coming from Wiktionary, LIWC, Wilson's subjectives, Hyland hedges, and Hooper's assertives, (\textit{iii})~writing style features such as TF.IDF-weighted character 3-grams, readability level and vocabulary richness (e.g.,~Flesch–Kincaid grade level, Flesch reading ease and the Gunning fog index), Type-Token Ratio (TTR), hapax legomena and dislegomena, and (\textit{iv})~the NEws LAndscape (NELA) features. The latter category includes 130 content-based features collected from the existing literature, which measure different aspects of a news article, comprising sentiment, bias, morality, and complexity~\cite{Horne2018}. The lexicon features are based on the analysis of the language of propaganda and trustworthy news, discussed in~\cite{rashkin-EtAl:2017:EMNLP2017}. The encouraging performance achieved by \texttt{Proppy} on reference datasets, as well as its capacity to outperform competing approaches and systems, makes it a favorable system to adopt for propaganda analysis.

\section{Analysis and Results}
\label{sec:results}
Below, we first analyse coordinated communities, we discuss our results and some limitations. We then combine these initial results with the analysis of propaganda, and we show how our combined approach helps to overcome the limitations of previous work.

\subsection{Finding Coordinated Communities}
\label{sec:results-coord}

The application of the method for investigating coordination to our dataset resulted in the user-similarity network shown in Figure~\ref{fig:network-communities}. The network is composed of seven communities of coordinated users, depicted with different colors in the figure and analytically described in Table~\ref{tab:communities} and in the rest of this subsection:

\noindent\tikz\draw[black, fill=lab] (0,0) circle (.75ex);~\texttt{LAB}: A large community of labourists that supported the Labour party and its leader Jeremy Corbyn, as well as traditional Labour themes such as healthcare and climate change.

\noindent\tikz\draw[black, fill=con] (0,0) circle (.75ex);~\texttt{CON}: A large community of conservative users. In addition to supporting the party and its leader Boris Johnson, this community was also strongly in favor of Brexit.

\noindent\tikz\draw[black, fill=tvt] (0,0) circle (.75ex);~\texttt{TVT}: A large community that included several parties, e.g.,~liberal democrats, who teamed up with labourists against the conservative party, a strategy dubbed \textit{tactical voting} in the 2019 UK election.

\noindent\tikz\draw[black, fill=snp] (0,0) circle (.75ex);~\texttt{SNP}: A medium-sized community of supporters of the Scottish National Party (SNP). These users also supported Scottish independence from the UK and asked for a new independence referendum.

\noindent\tikz\draw[black, fill=b60] (0,0) circle (.75ex);~\texttt{B60}: A small community of ``Backto60'' activitsts. Unlike the previous communities, these users did not represent a political party involved in the election. Instead, \texttt{B60} users leveraged the electoral debate to protest against a state pension age equalisation law that unfairly affected 4M women born in the 1950s.\footnote{\url{http://pensionsage.com/pa/Backto60-granted-leave-to-appeal.php}}

\noindent\tikz\draw[black, fill=ase] (0,0) circle (.75ex);~\texttt{ASE}: A small community of conservative users. Despite sharing the same political orientation, these users were separated from the users in the \texttt{CON} community because, rather than supporting the conservative party, they were mainly involved in attacking the labour party. An important narrative for \texttt{ASE} were antisemitism allegations targeted at labourists and Jeremy Corbyn throughout the electoral debate.\footnote{\url{http://www.thetimes.co.uk/article/revealed-the-depth-of-labour-anti-semitism-bb57h9pdz}}

\noindent\tikz\draw[black, fill=lch] (0,0) circle (.75ex);~\texttt{LCH}: Another small community of activists. Similarly to \texttt{B60}, these users were not particularly interested in the electoral debate, but rather protested against a retrospective taxation called ``loan charge'' that forced certain people to return unsustainable amounts, and which also resulted in several suicides.\footnote{\url{http://www.gov.uk/government/publications/disguised-remuneration-independent-loan-charge-review/guidance}}

The communities that emerged from the analysis of our user-similarity network are consistent with the 2019 UK political landscape and with the results of previous work~\cite{schumacher2019brexit,jackson2019uk}. Each of these communities had different goals, featured different narratives, and showed diverse degrees of coordination. 
In particular, we found both large and small communities, as shown in Figure~\ref{fig:network-communities}. While the larger communities are related to the main political parties in UK that participated in the election (e.g.,~\texttt{LAB}, \texttt{CON}, \texttt{TVT}, and \texttt{SNP}), the smaller ones represent other highly coordinated users who share a common goal, such as protesting activists (e.g.,~\texttt{B60} and \texttt{LCH}) and political antagonists (e.g.,~\texttt{ASE}). 
The analysis of Table~\ref{tab:communities} also reveals a few differences in the sharing behaviors of the different communities. Overall, larger communities seem to share less original articles and tweets, compared to the smaller groups. Moreover, previous analysis showed that these communities are characterized by a large negative assortativity, meaning that influential users are mostly connected to ordinary ones, and vice-versa~\cite{nizzoli2021coordinated}.

\begin{figure}[t]
    \centering
    \includegraphics[width=1\columnwidth]{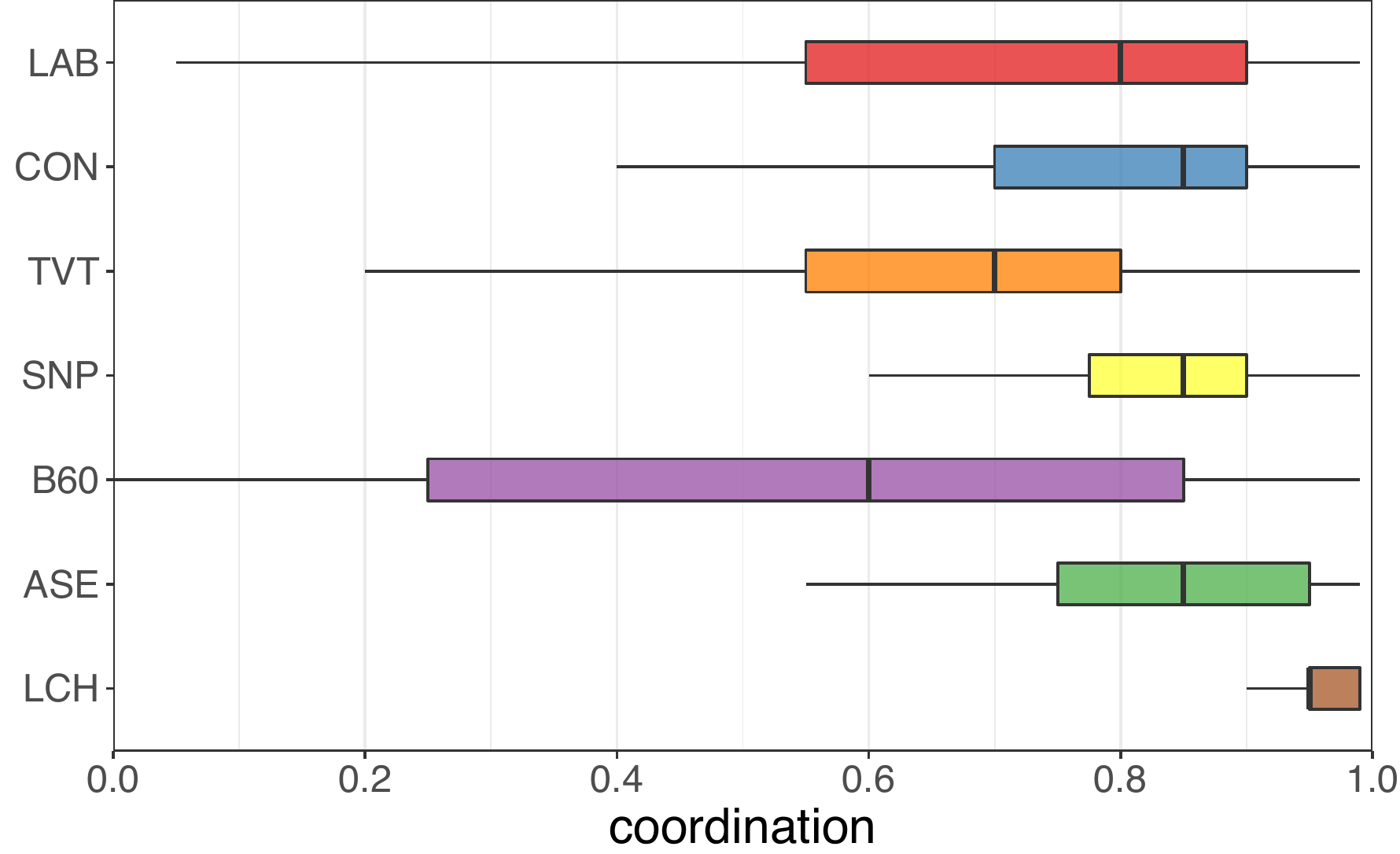}
    \caption{Distribution of coordination scores for users from the different communities involved in the online debate.}
    \label{fig:coordination-communities}
\end{figure}

These figures are indicative of top-down behaviors, where a small number of highly influential characters (e.g.,~the party leaders) drive the activities of the remaining members. In contrast, smaller communities seem to exhibit bottom-up behavior, characterized by grassroots activities and more content heterogeneity, as testified by the large percentage of original articles and tweets. 
The distribution of coordination scores for users of the different communities is shown in Figure~\ref{fig:coordination-communities}. Again, different behavior and characteristics emerge for the different communities, and particularly, for the smaller ones. For instance, while \texttt{B60} features users with diverse degrees of coordination, as shown by a relatively wide boxplot, \texttt{LCH} and \texttt{SNP} are much more homogeneous. \texttt{B60} is also the community with the lowest degree of average coordination, in contrast to \texttt{LCH} and, to a lower extent, to \texttt{ASE}, \texttt{SNP}, and \texttt{CON}.

\textbf{Discussion.} Our results so far match the state-of-the-art in the analysis of coordinated behavior~\cite{nizzoli2021coordinated,pacheco2020uncovering,weber2020s}. On the one hand, this approach allows us to obtain nuanced results in terms of coordinated communities. In fact, it allows us to detect several groups of coordinated users, both large and small, thus yielding more informative results compared to coarser analysis that only focused on the two main factions involved in an online debate (e.g.,~right vs. left, Democrats vs. Republicans, etc.), as done in~\cite{conover2011political,garimella2017reducing}. Moreover, it surfaces different patterns of coordination in the network (e.g.,~top-down vs. bottom-up). This approach to the analysis of coordinated behavior allows us to obtain nuanced and fine-grained results, typical of studies that require a great deal of offline, manual investigation~\cite{starbird2019disinfo,assenmacher2020two}, while still retaining the advantages of large-scale, automated analysis.

On the other hand, these evaluation results do not give insights into the harmfulness of the respective  coordinated communities. In other words, it is still not possible to clearly identify which communities (if any) of those shown in Figure~\ref{fig:network-communities} exploited coordination for tampering with the 2019 UK electoral debate on Twitter, and which instead represent neutral or well-intentioned coordinated users. Our subsequent analysis below contributes to answering this question.

\subsection{Measuring Propaganda on Social Media}
\label{sec:results-prop}

Our dataset contains two sources of textual content that can potentially convey propaganda: (\textit{i}) articles and (\textit{ii}) tweets. Thus, the first choice for computing propaganda scores is which items to analyze: articles or tweets. All propaganda detection systems so far -- including \texttt{Proppy}, the one we use in our analysis -- were developed for the analysis of news articles~\cite{dasanmartino2020survey}. However, from Table~\ref{tab:communities}, we notice that our dataset features, on average, less than one original news article per user and about 32 original tweets per user. Thus, basing our propaganda scores on articles would result in sparse and unreliable estimations. Moreover, the original tweets are authored by the users themselves, unlike news articles, which are just reshared. For these reasons, tweets arguably represent a more direct and reliable input for estimating a user's propaganda. Nonetheless, we computed propaganda scores based on both articles and tweets, and we subsequently compared and validated each of them. The outcome of this comparison and validation allowed us to identify a suitable propaganda score to use in the remainder of our study.

For computing propaganda scores based on articles, we used \texttt{Proppy} with the same configuration proposed by its authors in~\cite{BARRONCEDENO20191849}. For tweets, we made adjustments to account for the inherent differences between news articles and tweets. Specifically, several machine learning features used in the textual classifiers are influenced by document length, and tweets are obviously much shorter than news articles. Thus, we did not classify single tweets, but we grouped the original tweets (i.e., without retweets) by the same author into chunks whose length was comparable to that of the articles used to train \texttt{Proppy}. The grouping merged tweets in chronological order, but we did not apply any filtering based on their textual content (e.g., topic). We carried out a validation of our propaganda estimations by manually inspecting a subset of the tweets classified by \texttt{Proppy}, which revealed meaningful and satisfactory classifications, and supported our approach for detecting propagandistic vs. non-propagandistic tweets.

\begin{figure*}[t]
    \centering
    \begin{subfigure}[t]{.25\textwidth}%
        \includegraphics[width=\textwidth]{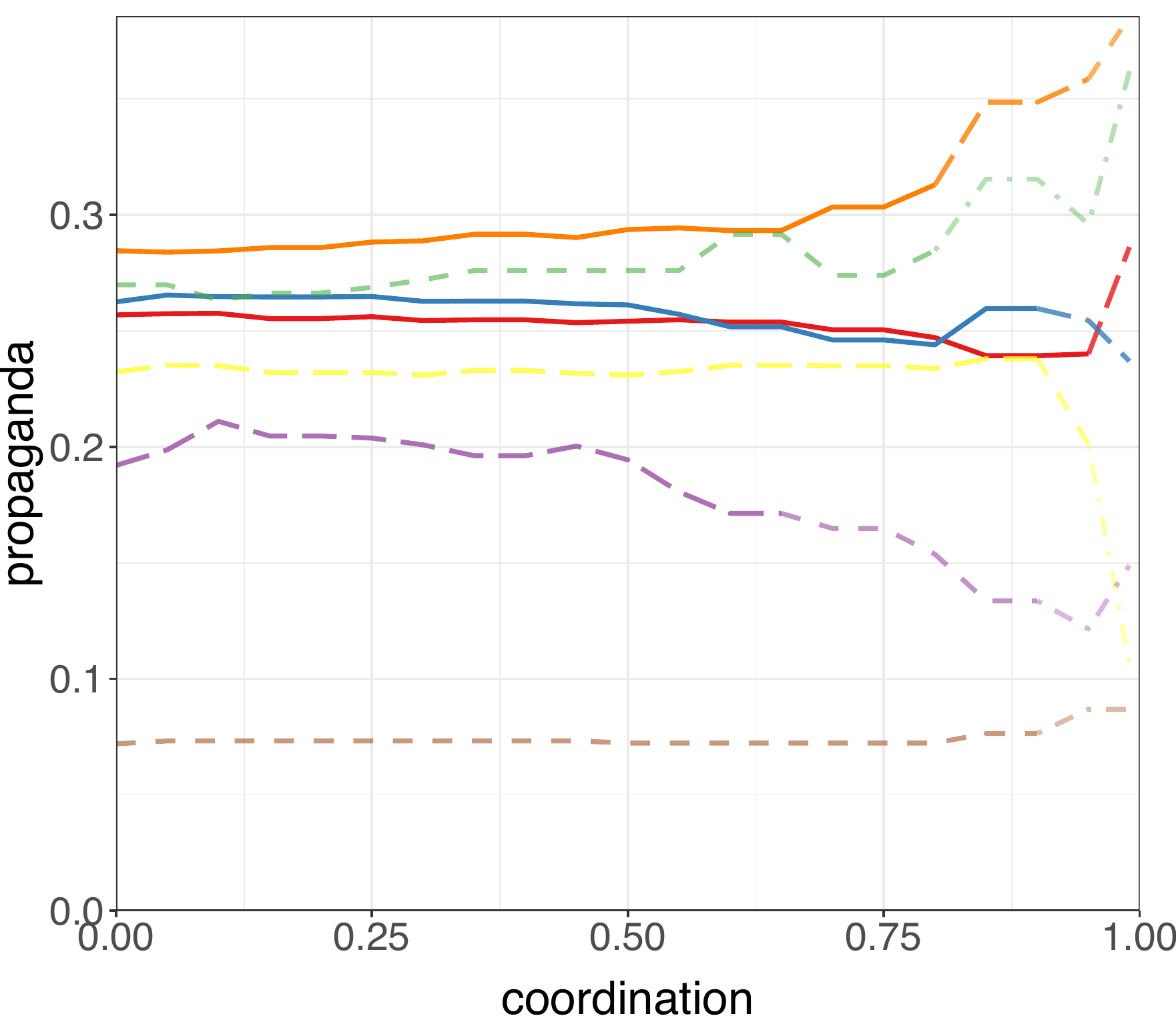}
        \caption{$M_1$.\label{fig:prop-metrics-tw-wmedian-mean}}
    \end{subfigure}%
    \begin{subfigure}[t]{.25\textwidth}%
        \includegraphics[width=\textwidth]{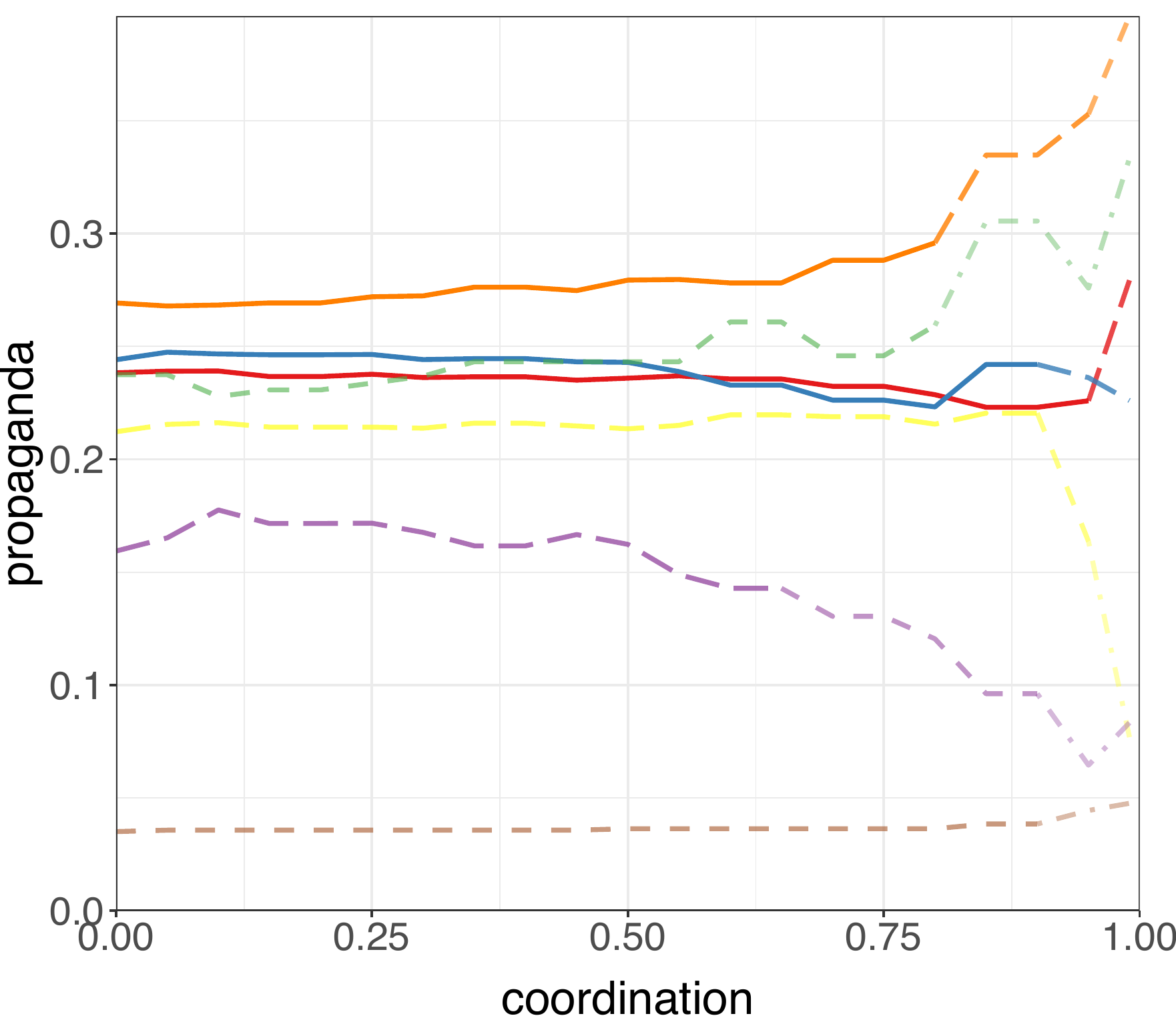}
        \caption{$M_2$.\label{fig:prop-metrics-tw-majvote-mean}}
    \end{subfigure}%
    \begin{subfigure}[t]{.25\textwidth}%
        \includegraphics[width=\textwidth]{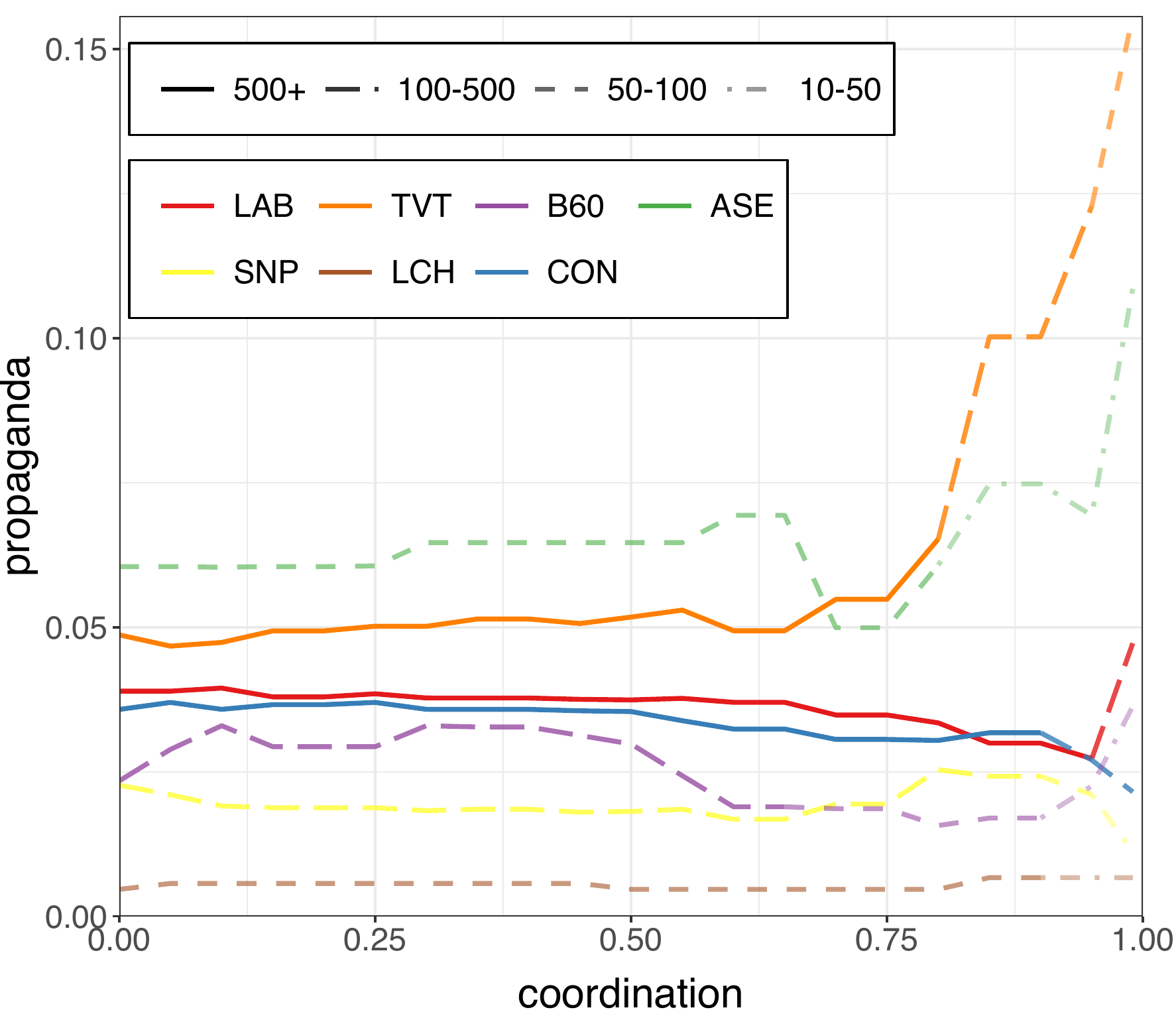}
        \caption{$M_{13}$.\label{fig:prop-metrics-tw-wmedian-median}}
    \end{subfigure}%
    \begin{subfigure}[t]{.25\textwidth}%
        \includegraphics[width=\textwidth]{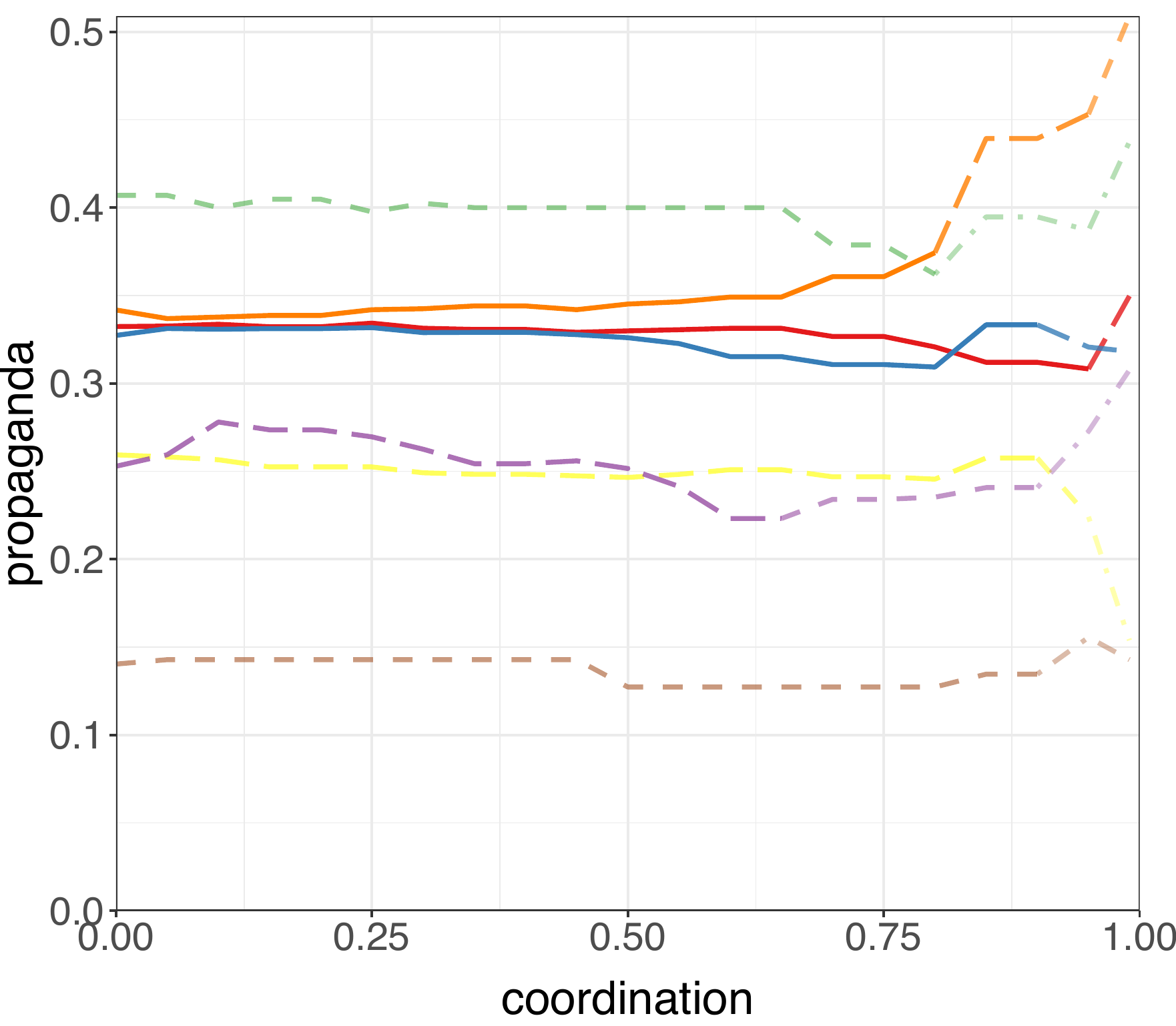}
        \caption{$M_{23}$.\label{fig:prop-metrics-tw-max-ratio}}
    \end{subfigure}%
    \caption{Examples of the relationship between propaganda and coordination for the different communities, obtained with a subset of the measures in Table~\ref{tab:informativeness}. Independently on the measure $M_i$, some communities consistently appear as more propagandistic than others. Specifically, \texttt{\textmd{TVT}} and \texttt{\textmd{ASE}} are among the communities that shared the most propaganda, while \texttt{\textmd{SNP}}, \texttt{\textmd{B60}} and \texttt{\textmd{LCH}} shared the least. Moreover, the most informative measures from Table~\ref{tab:informativeness}, i.e.,~$M_1$ and $M_2$, exhibit qualitatively similar propaganda trends.}
    \label{fig:prop-coord-per-community-differences}
\end{figure*}

Independently on the choice of analyzing articles or tweets, we obtained a propaganda score for user $j$ as follows: 
\[
P_u(u_j) = \Psi\big(P(i_k)\big) \; \forall \; i_k \; \text{shared by} \; u_j,
\]
where $\Psi$ is the user-level aggregation function, $i_k$ are all chunks of original tweets, or all distinct news articles, shared by $u_j$, and $P(i_k)$ are \texttt{Proppy}'s classifications of such items. 

Finally, since we want to compare different communities, we aggregate the user scores for each community. We compute the propaganda score of the $i$-th community $c_i$ as follows: 
\[
P_c(c_i) = \Phi\big(P_u(u_j)\big) \; \forall \; u_j \in c_i,
\]
where $\Phi$ is the community-level aggregation function. 

\begin{table}[tbh]
    \Small
    \centering
    \begin{tabular}{cclclr}
        \toprule
        & & \multicolumn{1}{c}{\textbf{user-level}} && \multicolumn{1}{c}{\textbf{community-level}} & \textbf{informativeness} \\
        \# & \textbf{item} & \multicolumn{1}{c}{\textbf{aggregation ($\Psi$)}} && \multicolumn{1}{c}{\textbf{aggregation ($\Phi$)}} & \textbf{($I$)} \\
        \midrule
        $M_1$ & \faTwitter & median && mean & 0.5491 \\ 
        $M_2$ & \faTwitter & majority voting && mean & 0.5457 \\
        $M_3$ & \faTwitter & majority voting && ratio & 0.5457 \\
        $M_4$ & \faTwitter & median && ratio & 0.5442 \\ 
        $\vdots$ & $\vdots$ & \multicolumn{1}{c}{$\vdots$} && \hspace{1em}$\vdots$ & $\vdots\hspace{1em}$ \\ [0.5em]
        $M_{10}$ & \faNewspaperO & median && mean & 0.5156 \\ 
        $M_{11}$ & \faNewspaperO & majority voting && mean & 0.4990 \\
        $M_{12}$ & \faNewspaperO & majority voting && ratio & 0.4990 \\
        $M_{13}$ & \faTwitter & median && median & 0.4955 \\ 
        $\vdots$ & $\vdots$ & \multicolumn{1}{c}{$\vdots$} && \hspace{1em}$\vdots$ & $\vdots\hspace{1em}$ \\ [0.5em]
        $M_{21}$ & \faNewspaperO & median && ratio & 0.4475 \\ 
        $M_{22}$ & \faTwitter & max && mean & 0.4400 \\
        $M_{23}$ & \faTwitter & max && ratio & 0.4400 \\
        $M_{24}$ & \faNewspaperO & median && median & 0.4214 \\ 
        \bottomrule
        \multicolumn{6}{l}{\faTwitter: tweets, \faNewspaperO~: articles}
    \end{tabular}
    \caption{Excerpt of the propaganda measures $M_i$ that we compared, shown in descending order of informativeness.\label{tab:informativeness}}
\end{table}

Different aggregation functions $\Psi$ and $\Phi$ (e.g., mean, median, max, etc.) can be used to compute $P_u$ and $P_c$, respectively. These, in addition to the choice of analyzing articles vs. tweets, result in many possible measures for computing propaganda scores. Table~\ref{tab:informativeness} lists some of the measures that we experimented with. In the next sections, we compare the informativeness of these measures, and we choose a suitable one for our further analysis.

\subsection{Spread of Propaganda by Coordinated Users}
\label{sec:results-communities}

\textbf{Combining coordination and propaganda.} So far, our approach provided us with three pieces of information that we can combine: (\textit{i})~communities, (\textit{ii})~coordination scores, and multiple (\textit{iii})~propaganda scores. By combining community labels with coordination and propaganda scores, we can study the trends of propaganda as a function of coordination for each community. Let $C_u(u_j)$ be the coordination score of the $j$-th user $u_j$. Then, the propaganda score for community $c_i$, as a function of coordination, is defined as
\begin{equation}
\label{eq:prop-coord}
    P_c(c_i, k) = \Phi\big(P_u(u_j)\big) \; \forall \; u_j \in c_i : C_u(u_j) \ge k.
\end{equation}
In other words, Equation~\eqref{eq:prop-coord} defines how to compute a propaganda score at different coordination thresholds $k$, for each community. Therefore, we can assess whether the most coordinated users in each community were also the most propagandistic ones. In turn, this provides valuable information for assessing the harmfulness (or lack thereof) of the different communities.

Figure~\ref{fig:prop-coord-per-community-differences} shows examples of this analysis obtained by applying Equation~\eqref{eq:prop-coord} at different levels of coordination, for some of the propaganda measures shown in Table~\ref{tab:informativeness}. In this figure, we use different line types and transparencies to indicate the number of users in each community at different levels of coordination. In fact, each community has a different cardinality, as reported in Table~\ref{tab:communities}. Moreover, fewer users are considered when moving towards large coordination values, i.e.,~only the most coordinated ones. As an example, Figure~\ref{fig:prop-metrics-tw-wmedian-mean} shows that, for each community, we always have more than 10 users (as well as the several hundreds of tweets that they shared), even at coordination of about 1. Moreover, we always have more than 50 users and thousands of tweets, when the coordination is 0.8 or lower. For propaganda scores derived from tweets, this ensures that the trends shown in the figure are not derived from a trivial number of tweets.

\begin{table*}[t]
    \small
    \centering
    \begin{tabular}{clcrlcrlcrlcrl}
        \toprule
        \multicolumn{2}{l}{\textbf{community}} && \multicolumn{2}{c}{\textbf{coordination}} && \multicolumn{2}{c}{\textbf{automation}} && \multicolumn{2}{c}{\textbf{suspensions}} && \multicolumn{2}{c}{$\delta$ (\%)} \\
        \midrule
        \tikz\draw[black, fill=lab] (0,0) circle (.75ex); & \texttt{LAB} && $-0.193$ &  && $0.428$ & * && $0.925$ & *** && $-0.016$ & ($-6.5\%$) \\
        \tikz\draw[black, fill=con] (0,0) circle (.75ex); & \texttt{CON} && $-0.754$ & *** && $-0.688$ & *** && $-0.079$ &  && $-0.008$ & ($-3.0\%$) \\
        \tikz\draw[black, fill=tvt] (0,0) circle (.75ex); & \texttt{TVT} && $0.813$ & *** && $0.836$ & *** && $0.844$ & *** && $+0.074$ & ($+26.0\%$) \\
        \tikz\draw[black, fill=snp] (0,0) circle (.75ex); & \texttt{SNP} && $-0.404$ & * && $0.358$ &  && $0.902$ & *** && $-0.031$ & ($-13.5\%$) \\
        \tikz\draw[black, fill=b60] (0,0) circle (.75ex); & \texttt{B60} && $-0.899$ & *** && $-0.961$ & *** && $0.882$ & *** && $-0.070$ & ($-36.7\%$) \\
        \tikz\draw[black, fill=ase] (0,0) circle (.75ex); & \texttt{ASE} && $0.742$ & *** && $0.762$ & *** && $-0.218$ &  && $+0.026$ & ($+9.7\%$) \\
        \tikz\draw[black, fill=lch] (0,0) circle (.75ex); & \texttt{LCH} && $0.562$ & *** && $0.704$ & *** && $-0.733$ & *** && $+0.014$ & ($+20.7\%$) \\ [0.75em]
        \multicolumn{2}{l}{overall} && $-0.679$ & *** && $-0.285$ &  && $-0.052$ &  && \multicolumn{2}{c}{--} \\
        \midrule
        &&& \multicolumn{2}{c}{(a)} && \multicolumn{2}{c}{(b)} && \multicolumn{2}{c}{(c)} &&& \\
        \bottomrule
        \multicolumn{14}{l}{***: $p < 0.01$, **: $p < 0.05$, *: $p < 0.1$}
    \end{tabular}
    \caption{Correlation analysis between propaganda scores and (a)~coordination scores, (b)~automation scores, and (c)~Twitter suspensions (for all communities).\label{tab:corr-prop-vs-all}}
\vspace*{-10pt}
\end{table*}

\textbf{Choosing a suitable propaganda measure.} Below, we report the results of a qualitative comparison and a quantitative evaluation of the propaganda measures from Table~\ref{tab:informativeness}. A desirable characteristic of a propaganda measure is the capacity to distinguish propagandistic vs. non-propagandistic communities. That is, its capacity to highlight the differences between the several communities involved in the online electoral debate, with respect to the use of propaganda. 

We evaluated the \textit{informativeness} ($I$) of each measure based on the differences between the propaganda trends that it produces. We quantified the difference between the propaganda trends for two communities as the possible negative linear correlation between them. Thus, for any given measure, we computed the average of the Pearson's correlations $r$ between the propaganda trends $P_c(c_x, k)$ and $P_c(c_y, k)$ of each possible pair of communities $c_x$ and $c_y$: 
\[
\Bar{r} = \frac{1}{N}\sum_{x,y=1}^N{r\big(P_c(c_x, k), P_c(c_y, k)\big)} \; \forall \; x,y : x \ne y.
\]
Then, we computed the informativeness of a measure as $I = \frac{1-\Bar{r}}{2}$. 

Intuitively, if a measure yields a \textit{positive} correlation between propaganda community trends, then $\Bar{r} \approx 1$ and $I \approx 0$. This means that such a measure is not able to diversify the behavior of the different communities, as reflected by the low informativeness. 

Conversely, if a measure yields a large \textit{negative} correlation between propaganda community trends, then $\Bar{r} \approx -1$ and $I \approx 1$, meaning that the measure can diversify the different communities. Notably, our approach is similar but favorable over other alternatives for measuring informativeness, such as those based on mutual information, since they require additional problematic steps for estimating unknown distributions, as discussed in~\cite{khan2007relative}.


\begin{figure*}[t]
    \centering
    \begin{subfigure}[t]{.25\textwidth}%
        \includegraphics[width=\textwidth]{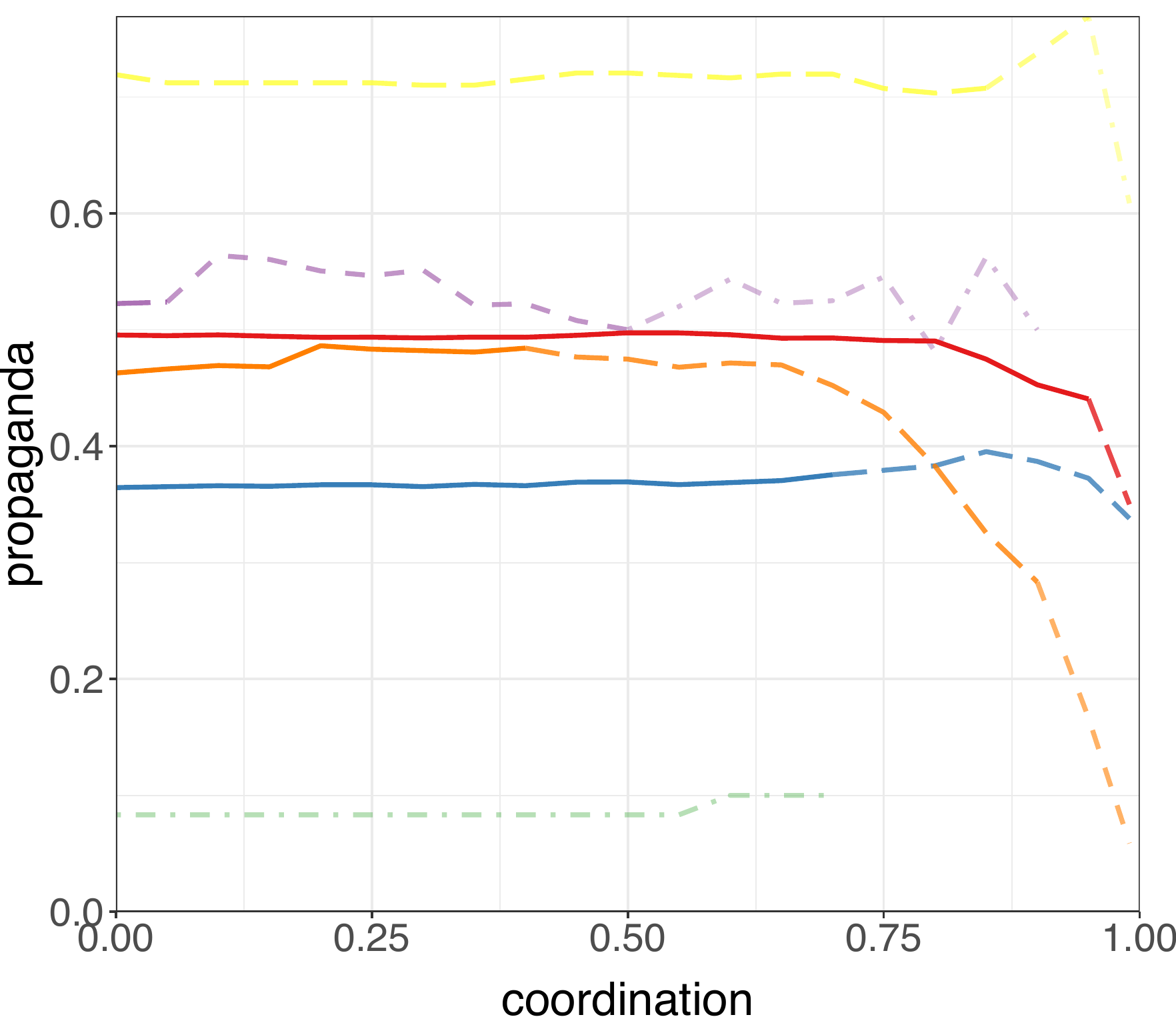}
        \caption{policy prescription\label{fig:policy-prescription-framing}}
    \end{subfigure}%
    \begin{subfigure}[t]{.25\textwidth}%
        \includegraphics[width=\textwidth]{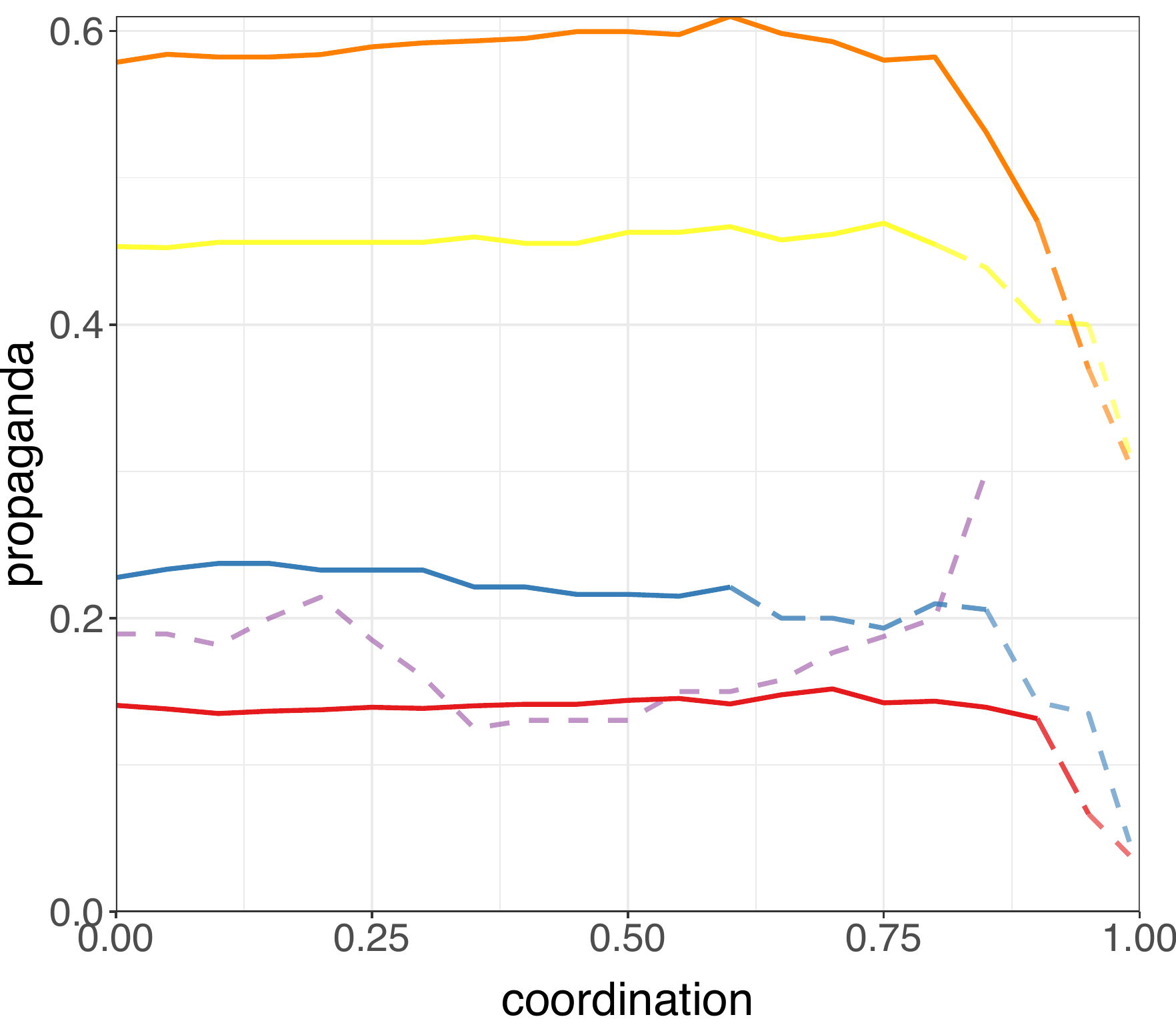}
        \caption{public opinion\label{fig:public-opinion-framing}}
    \end{subfigure}%
    \begin{subfigure}[t]{.25\textwidth}%
        \includegraphics[width=\textwidth]{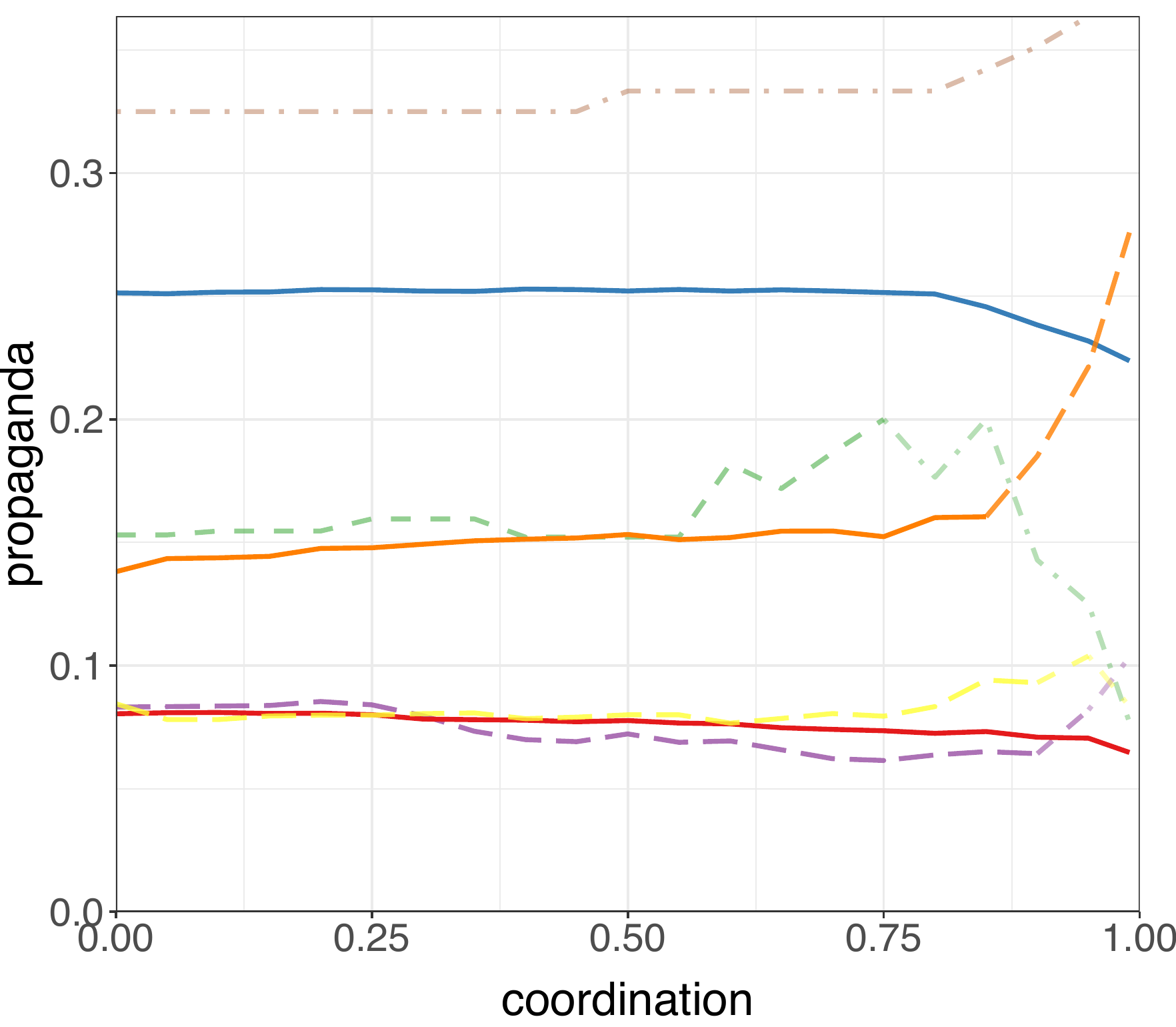}
        \caption{economy\label{fig:economic-framing}}
    \end{subfigure}%
    \begin{subfigure}[t]{.25\textwidth}%
        \includegraphics[width=\textwidth]{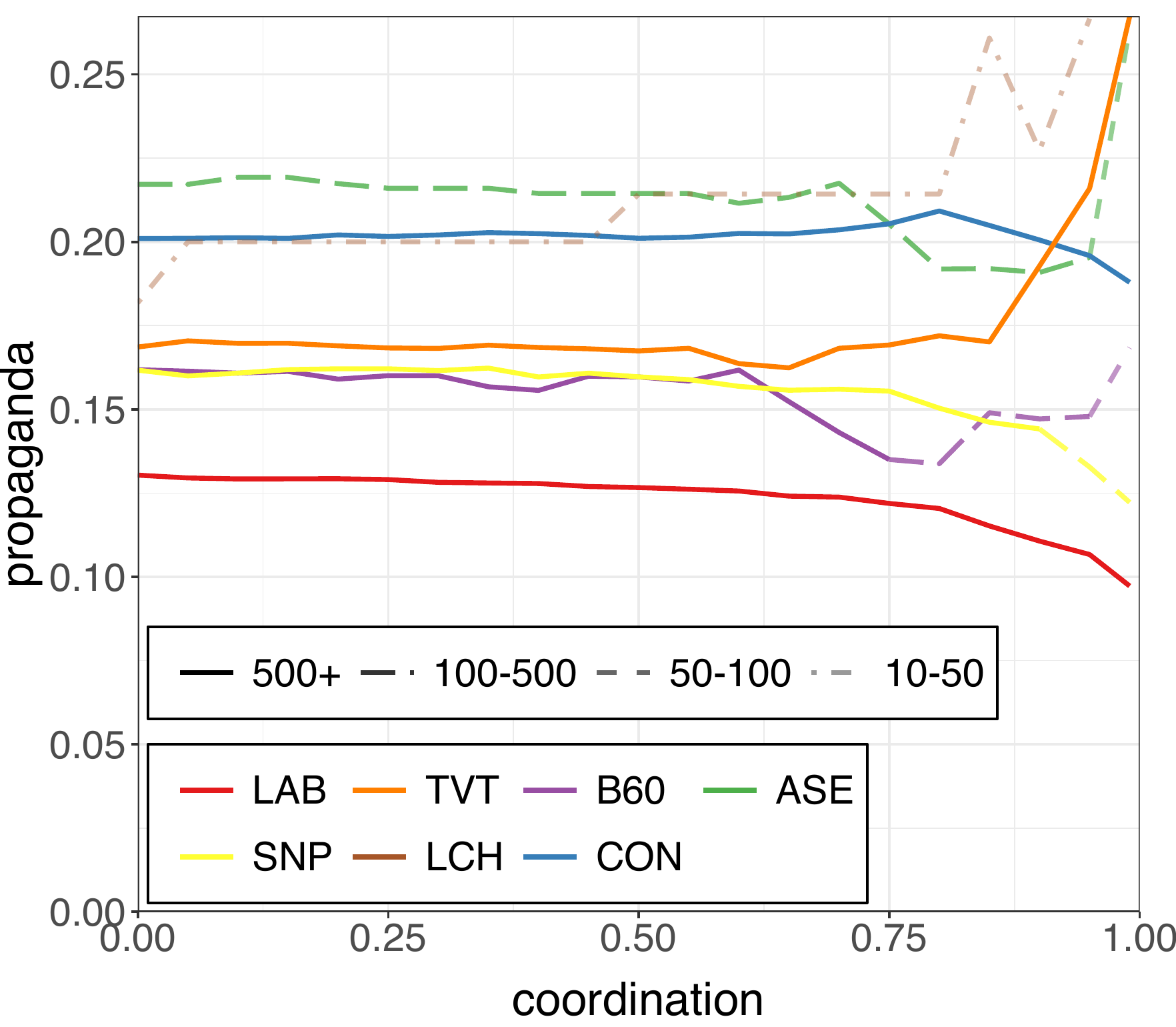}
        \caption{political\label{fig:political-framing}}
    \end{subfigure}%
    \caption{\label{fig:propaganda-coord-per-community-framing}Frame analysis of the news articles shared by the different communities. Propaganda is measured as the fraction of articles in a certain frame flagged as propagandist, out of all articles from that frame. Line-types indicate the number of tweets. Lines may interrupt if no tweets were shared by coordinated users (coordination $\geq k$) of a community for a given frame.}
\end{figure*}

The last column of Table~\ref{tab:informativeness} reports the informativeness of each propaganda measure, with $M_1$ being the most informative one. As shown in the table, there are relatively small differences in the informativeness of the propaganda measures that we evaluated, which ranges between 0.42 and 0.55 on a 0 to 1 scale. This means that the majority of the measures yield comparable results, as is also visible by the qualitatively similar propaganda trends shown in Figures~\ref{fig:prop-metrics-tw-wmedian-mean} and~\ref{fig:prop-metrics-tw-majvote-mean}. This suggests that changing the measure would not drastically alter the evaluation results. Nonetheless, the topmost measures in Table~\ref{tab:informativeness} are relatively better (i.e.,~more informative) at surfacing the differences between the investigated communities. Interestingly, the most informative measures in Table~\ref{tab:informativeness} are all based on analysis of tweets, which reinforces our initial hypothesis and the results of our manual validation of propaganda classifications of tweets. Following these preliminary results and without loss of generality, in our subsequent analysis we adopt $M_1$ for measuring the propaganda scores. 


\textbf{Results.} 
Figure~\ref{fig:prop-metrics-tw-wmedian-mean} shows interesting propaganda trends for some communities. First, \texttt{LCH} is characterized by the lowest degree of propaganda among all communities. Similarly, \texttt{B60} shows a marked decreasing propaganda trend, implying that the core users of the community (i.e.,~the most coordinated ones) are not engaged in propaganda. Both these findings suggest harmless behavior. In other words, \texttt{LCH} appears to be highly coordinated, as shown in Figure~\ref{fig:coordination-communities}, but harmless. \texttt{B60} features diverse degrees of coordination among its users, but is nonetheless harmless. On the contrary, other communities feature increasing propaganda trends: above all, \texttt{TVT} and \texttt{ASE}. For coordination of 0.5 and above, both show increasing levels of propaganda, which supports the hypothesis of harmful communities. For the remaining three communities (i.e.,~\texttt{LAB}, \texttt{CON}, and \texttt{SNP}) the coordination appears to be mostly unrelated to propaganda. 

Note that these results are overall robust to changing the measure used to compute the propaganda scores.

The above qualitative findings are further confirmed by the quantitative results reported in Table~\ref{tab:corr-prop-vs-all}a. In particular, our correlation analysis shows strong, positive, and statistically significant Pearson correlations between propaganda and coordination for the \texttt{TVT} and the \texttt{ASE} communities, with $r = 0.813$ and $r = 0.742$, respectively. Instead, \texttt{B60} features a strong, negative, and statistically significant correlation $r = -0.899$. The remaining results in Table~\ref{tab:corr-prop-vs-all} are not meaningful, either because of small correlations or due to low statistical significance (for \texttt{LAB}, \texttt{SNP}, and \texttt{LCH}), or because of limited variation of propaganda (for \texttt{CON}). Regarding the latter, Pearson correlation measures the strength of the linear relationship between two variables, but not the extent of variation for either one, which is relevant in our analysis. The last column of Table~\ref{tab:corr-prop-vs-all} accounts for this aspect by measuring the variation in propaganda as $\delta = P_c(c_i, k = 0.9) - P_c(c_i, k = 0)$, for each community $c_i$. Despite featuring a marked negative and significant correlation, \texttt{CON} exhibits a very small variation in propaganda, with $\delta = -0.008$, represented in Figure~\ref{fig:prop-metrics-tw-wmedian-mean} by a mostly flat line.

\begin{table*}[t]
    \scriptsize
    \centering
    \begin{tabular}{cp{0.6\linewidth}p{0.32\linewidth}}
        \toprule
        & \textbf{tweet text} & \textbf{article title} \\
        \midrule
        \multirow{11}{*}{\texttt{TVT}} & Join @fckboris \& \#RegisterToVote \#GE2019 \textit{$<$url$>$} & Five ways to say Fck Boris \\ \cline{2-3} 
        & GE2019: Five ways to say F**k Boris \#GeneralElection2019\#Brexit\#VoteToriesOut \textit{$<$url$>$}  & Five ways to say Fck Boris  \\ \cline{2-3} 
        & KARMA: Katie Hopkins forced to sell £1m home and now rents after cringe libel loss / \#GE2019 \#Brexit \textit{$<$url$>$} & Katie Hopkins forced to sell £1m home and now rents after humiliating libel case \\ \cline{2-3} 
        & Here is vocal \#RemainerNow and life long Tory voter @OborneTweets on \#GE2019 \textit{$<$url$>$} & As a lifelong Conservative, here’s why I can’t vote for Boris Johnson \\ \cline{2-3} 
        & Boris Johnson unfit to be Prime Minister. Brexit damages the Henley constituency -  John Howell MP knows this but continues his support-  @LauracoyleLD Laura Coyle the People’s Vote recommendation \textit{$<$url$>$} \#TacticalVoting \#PeoplesVote \#Boris \#GeneralElection2019 & I was Boris Johnson’s boss: he is utterly unfit to be prime minister \\ \cline{2-3} 
        & This is beyond desperate and beyond tin pot banana republic actions if true. \#GE2019 \textit{$<$url$>$} & General election: Farage claims No 10 offered Brexit Party candidates jobs to stand down \\
        \midrule
        \multirow{11}{*}{\texttt{B60}} & I’m voting Labour for the final say on Brexit. Share why you’re voting Labour Right \#VoteLabour \textit{$<$url$>$} & I'm Voting Labour… \\ \cline{2-3} 
        & I’m voting Labour for a million green jobs. Share why you’re voting Labour \#VoteLabour  \textit{$<$url$>$}  & I'm Voting Labour…  \\ \cline{2-3}
        & WASPI Women won’t be silenced \#GeneralElection2019 \textit{$<$url$>$} via @WASPI\_Campaign & WASPI Women won’t be silenced \#GeneralElection2019 \\ \cline{2-3}
        & Owen Jones: ‘They don’t want you to vote. Defy them’ \#GTTO  \#GeneralElection2019 \textit{$<$url$>$} Sent via @updayUK & Owen Jones: ‘They don’t want you to vote. Defy them’  \\ \cline{2-3} 
        & Well done May from Falkirk Waspi. \textit{$<$url$>$} \#waspicampaign2018 \#deedsnotwords \#onevoice \#GeneralElection2019 & WASPI woman puts Boris Johnson on spot about trust after he publicly pledged to try and sort out pension row during visit to Cheltenham \\ \cline{2-3}
        & Now that most party manifestos have been published, join  @WASPI\_Campaign today at \textit{$<$url$>$} and find your nearest local group at \textit{$<$url$>$} to find out about our \#GE2019 Toolkit so you can speak up for \#WASPI in your local area \#WASPIwomenvote & How to join WASPI \\ 
        \bottomrule
    \end{tabular}
\caption{\label{tab:tweet-examples}Excerpt of the activity of strongly coordinated ($k \geq 0.9$) members of \texttt{\textmd{TVT}} and \texttt{\textmd{B60}}. While the \texttt{\textmd{TVT}} users attack Boris Johnson and Brexit, the \texttt{\textmd{B60}} users encourage women to vote and support the WASPI (Women Against State Pension Inequality) campaign.\protect\footnotemark \ Our method labeled all \texttt{\textmd{TVT}} tweets in the table as propagandistic, and all \texttt{\textmd{B60}} tweets as non-propagandistic.} 
\end{table*}

\textbf{Discussion.} Given the lack of ground truth on coordinated harmful vs. harmless behaviors, one way to qualitatively validate our analysis is by cross-checking our results with previous work and with the role of the communities in the electoral debate, as also done in previous work~\cite{nizzoli2021coordinated,pacheco2020uncovering}. Two communities emerged as authentic and harmless: \texttt{LCH} and \texttt{B60}. This means that, according to our proposed methodology, their activities are coordinated, but not malicious nor deceptive. In other words, they exhibit coordinated but authentic and harmless behavior. From previous work~\cite{nizzoli2021coordinated} and from our analysis of coordinated communities, we find that these are groups of activists protesting against unfair taxation (\texttt{LCH}) and in favor of women's rights (\texttt{B60}). Table~\ref{tab:tweet-examples} provides a detailed look at some of the tweets from \texttt{B60}'s highly coordinated users, confirming that their focus was to promote their cause and to encourage women to exercise their right to vote. They further endorsed the Labour leader Jeremy Corbyn, who expressed support for their initiative. Hence, our methodology correctly highlighted these activists as harmless examples of grassroots coordination. In contrast, our analysis revealed that \texttt{TVT} and \texttt{ASE} featured characteristics related to harmful behaviors. Both are highly polarized communities with strong political motivations.

Regarding \texttt{TVT} and its highly coordinated members, Table~\ref{tab:tweet-examples} shows that the majority of their tweets and shared articles are politically themed. Most of the time they attack Boris Johnson and the conservatives. Similarly, at the opposite side of the political spectrum, \texttt{ASE}'s peculiarity was that of repeatedly attacking the Labour party and its leader with allegations of antisemitism. Here, our methodology highlighted aggressive communities as harmful. Lastly, \texttt{LAB}, \texttt{CON}, and \texttt{SNP} appear as neither markedly harmless nor harmful. This is in line with the role of these communities, since they are large communities of moderate users~\cite{nizzoli2021coordinated}.

\footnotetext{\url{https://www.waspi.co.uk/about-us-2/}}

We further inspected the framing of the articles shared by different communities. We used the frame inventory of the Media Frames Corpus~\cite{card-etal-2015-media}, and we performed automatic annotation of the frames using the Tanbih API.\footnote{\url{http://app.swaggerhub.com/apis/yifan2019/Tanbih/0.8.0\#/}} Tanbih is a news aggregator platform with intelligent data analysis capabilities, including the possibility to analyze articles and news outlets based on their degree of factual reporting, propagandistic content, hyper-partisanship, political bias, and stance with respect to various claims and topics~\cite{zhang2019tanbih,darwish2020unsupervised,fagni2022fine}. Figure~\ref{fig:propaganda-coord-per-community-framing} shows the analysis for four relevant frames, highlighting striking differences across the frames, even within a single community. For example, Figure~\ref{fig:political-framing} shows that the \emph{Political} frame for \texttt{TVT} evolves into propagandistic behavior as coordination increases. Conversely, for \emph{Policy Prescription}, the propaganda score for the same community decreases. This suggests that the spread of propaganda is a theme-dependent phenomenon. 
Figure~\ref{fig:propaganda-coord-per-community-framing} also highlights that some communities deviate from the rest in terms of overall propagandistic content, such as~\texttt{SNP} and \texttt{TVT}, which maintain a relatively high propaganda score of 0.45--0.60 for \emph{Public Opinion}, compared to the remaining communities.

\textbf{Comparisons.} Here, we discuss our methodology and our results compared to previous work, highlighting the usefulness and the advantages of our approach. Several earlier attempts at detecting inauthentic and harmful campaigns only investigated coordination and synchronization between user accounts~\cite{pacheco2020uncovering,sharma2021identifying,weber2020s}. In this work, all groups of users exhibiting unexpected coordination were considered to be malicious~\cite{pacheco2020uncovering}. Despite representing an initial solution to the task of detecting malicious campaigns, this approach has a number of drawbacks. 

For example, if applied to our dataset, it would have flagged the \texttt{LCH} community as malicious, due to its extreme degree of coordination, as can be seen in Figure~\ref{fig:coordination-communities}. However, our nuanced analysis of propaganda and coordination revealed that the \texttt{LCH} users are protesting activists, which is a finding also confirmed by~\citet{nizzoli2021coordinated}. Conversely, the \texttt{TVT} community features the second-lowest degree of coordination among our communities. As such, it would have been labeled as non-suspicious by previous techniques. Our analysis further revealed a strong positive correlation between propaganda and coordination for \texttt{TVT} users, thus uncovering their malicious intent. In summary, our results show that coordination alone does not provide enough information for assessing the real activities and intent of online communities. Instead, a methodology combining the analysis of coordination with signs of malicious intent (e.g.,~propaganda), such as the proposed one, can distinguish inauthentic and harmful behavior vs. authentic and harmless one.

Our findings also confirm and extend previous results about the role of small and fringe Web communities in information disorder. \citet{zannettou2017web,zannettou2018origins} noted that fringe, polarized, and strongly motivated communities are those that exert the most influence on the Web regarding issues such as disinformation and online abuse, despite being relatively small. In our analysis, we obtained comparable results. Indeed, the most interesting communities (i.e.,~those that exhibit coordinated yet markedly harmless or harmful behavior) are small and non-mainstream, such as \texttt{LCH}, \texttt{B60}, and \texttt{ASE}. However, while \citeauthor{zannettou2018origins} investigated this phenomenon \textit{across} Web platforms, here we show that the same also occurs \textit{within} platforms.


\subsection{Inauthentic and Harmful Behavior}
\label{sec:results-inauthentic}

\begin{figure}[t]
    \centering
    \begin{subfigure}[t]{.5\columnwidth}%
        \includegraphics[width=\textwidth]{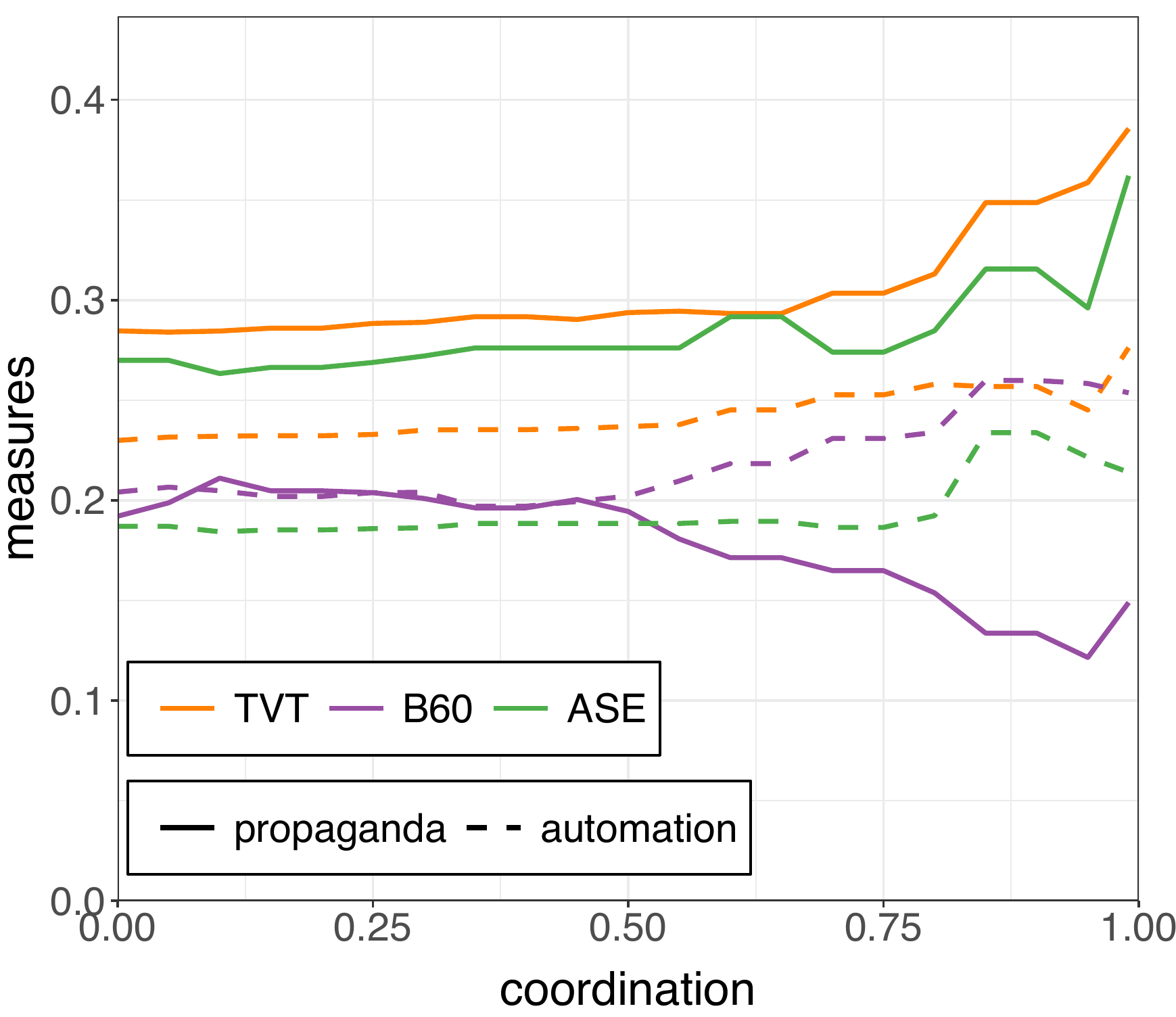}
        \caption{Propaganda vs. automation.\label{fig:corr-prop-vs-automation}}
    \end{subfigure}%
    \begin{subfigure}[t]{.5\columnwidth}%
        \includegraphics[width=\textwidth]{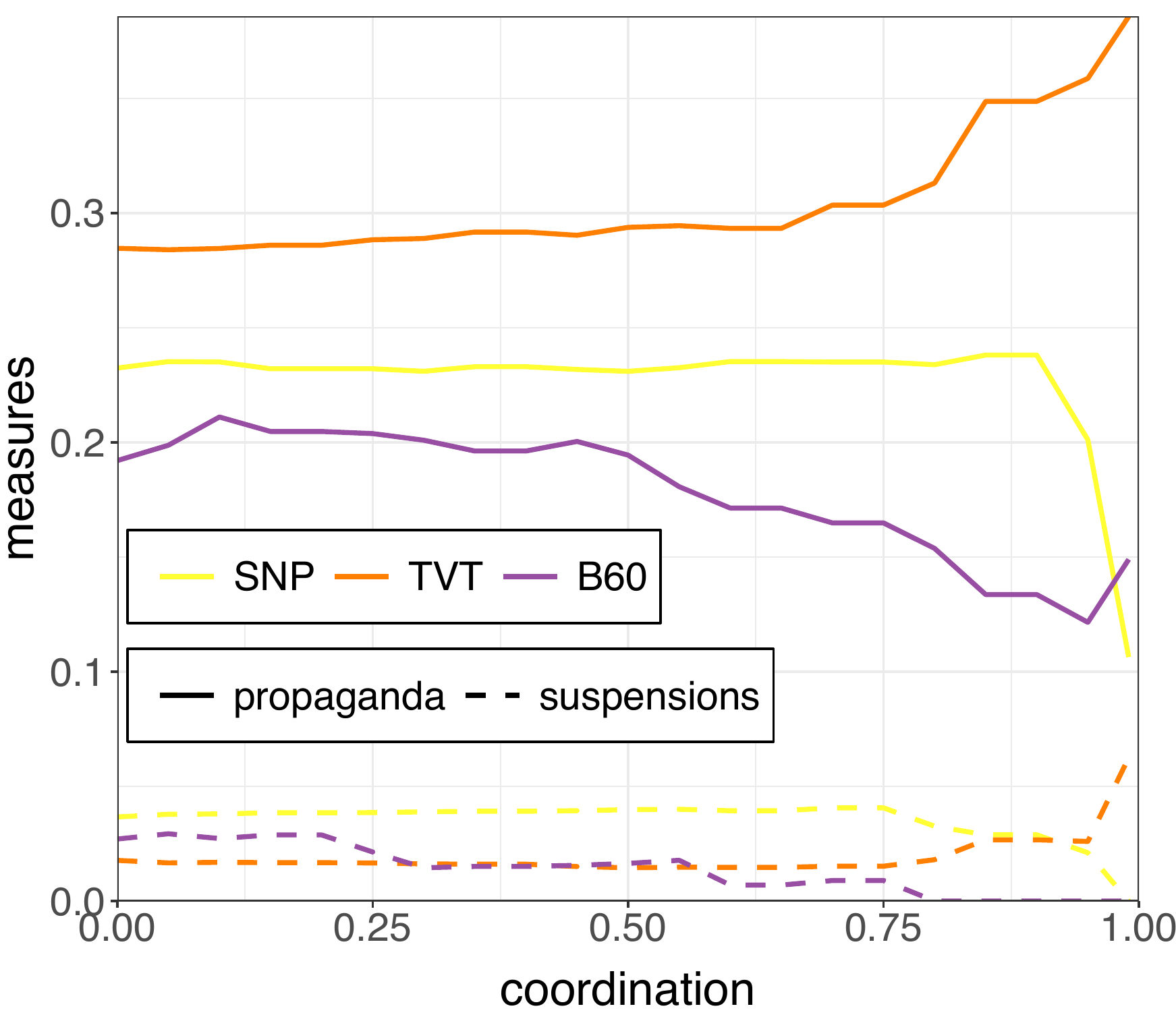}
        \caption{Propaganda vs. suspensions.\label{fig:corr-prop-vs-suspensions}}
    \end{subfigure}%
    \caption{Trends of propaganda versus trends of automation scores and suspensions, for a few relevant communities.}
    \label{fig:corr-prop-vs-misbehavior}
\end{figure}

We conclude our analysis by comparing propaganda and coordination scores to other clear signs of inauthenticity and harmfulness. We leverage scores indicating automation (i.e.,~botness) as a proxy for inauthenticity. For each account, we use the maximum of Botometer's English and universal scores, both provided in the $[0, 1]$ range, as its automation score~\cite{sayyadiharikandeh2020detection}. While we are aware of the limitations of current bot detectors~\cite{cresci2020decade}, including Botometer~\cite{rauchfleisch2020false}, the strong interest on the role played by social bots in online manipulation campaigns motivates this analysis. Similarly, we investigate the number of accounts suspended by Twitter in each community, as a proxy for harmfulness. Table~\ref{tab:corr-prop-vs-all} reports in columns (b) and (c) the correlation results between our propaganda scores versus automation and Twitter suspensions, respectively. By cross-checking strong and significant correlations against notable variations in propaganda ($\delta$), we highlight interesting trends. Regarding automation, the same communities that featured a strong positive correlation between propaganda and coordination -- namely, \texttt{TVT} and \texttt{ASE} -- are also strongly correlated with automation scores. This means that highly coordinated users in \texttt{TVT} and \texttt{ASE} are both inauthentic and harmful, further confirming our earlier results. An unexpected result is instead obtained for \texttt{B60}, which features a strong negative correlation between propaganda and automation. In other words, while propaganda decreases as a function of coordination, automation scores increase. Thus, coordinated \texttt{B60} users could be leveraging automation as a way to boost their online actions. Propaganda and automation trends for \texttt{TVT}, \texttt{ASE}, and \texttt{B60} are shown in Figure~\ref{fig:corr-prop-vs-automation}.

Overall, \texttt{TVT} appears as the most harmful community throughout the online UK electoral debate, with high propaganda, high automation, and the largest share of accounts suspended by Twitter. We further measure strong positive correlation between propaganda and suspension trends for \texttt{B60} and \texttt{SNP}. Since, for these communities, propaganda decreases with coordination, these positive correlations mean that Twitter suspensions also decrease, which is a sign of harmless behavior. This result is particularly relevant for \texttt{B60}, and it corroborates our previous findings. The trends of propaganda and suspensions for \texttt{SNP}, \texttt{TVT}, and \texttt{B60} are shown in Figure~\ref{fig:corr-prop-vs-suspensions}.

\section{Conclusion and Future Work}
\label{sec:conclusions}

We carried out the first combined analysis of propaganda and coordination in online debates. Specifically, we applied our methodology to the 2019 UK electoral debate on Twitter, revealing (\textit{i})~harmful, (\textit{ii})~neutral, and (\textit{iii})~well-intentioned communities that took part in the debate. Among the most harmful communities, we found ``tactical voters'' (\texttt{TVT}), who colluded against conservatives, and a small group of political antagonists, who attacked labourists and Jeremy Corbyn with accusations of antisemitism (\texttt{ASE}). Among the harmless coordinated communities, we uncovered groups of activists protesting against loan taxation (\texttt{LCH}) and in favor of women's rights (\texttt{B60}). Besides providing novel and interesting insights into the communities that participated in the 2019 UK electoral debate, our results also demonstrate the need to combine analysis of coordinated user behavior and intent. Our methodology contributes to distinguishing between coordinated \textit{harmful} and \textit{harmless} behavior, thus overcoming one of the main limitations of earlier work.

Among the future challenges along this important research direction is the construction of a reliable ground truth for coordinated harmful and harmless behavior. This endeavor would allow shifting from the current descriptive work to predictions, by training models that can detect harmful behavior. We also plan to collect and to investigate additional information about online communities, thus going beyond the analysis of coordination and propaganda. 
If successful, these efforts will allow a deeper understanding of coordinated online behavior, thus enabling the possibility to rapidly intervene, and ultimately to limit the spread, the influence, and the societal impact of online information disorders.

\begin{acks}
The work is part of the Tanbih mega-project, developed at the Qatar Computing Research Institute, HBKU, which aims to limit the impact of ``fake news,'' propaganda, and media bias by making users aware of what they are reading, thus promoting media literacy and critical thinking.
\end{acks}


\balance
\bibliographystyle{ACM-Reference-Format}
\bibliography{bibliography.bib}


\begin{thebibliography}{50}


\ifx \showCODEN    \undefined \def \showCODEN     #1{\unskip}     \fi
\ifx \showDOI      \undefined \def \showDOI       #1{#1}\fi
\ifx \showISBNx    \undefined \def \showISBNx     #1{\unskip}     \fi
\ifx \showISBNxiii \undefined \def \showISBNxiii  #1{\unskip}     \fi
\ifx \showISSN     \undefined \def \showISSN      #1{\unskip}     \fi
\ifx \showLCCN     \undefined \def \showLCCN      #1{\unskip}     \fi
\ifx \shownote     \undefined \def \shownote      #1{#1}          \fi
\ifx \showarticletitle \undefined \def \showarticletitle #1{#1}   \fi
\ifx \showURL      \undefined \def \showURL       {\relax}        \fi
\providecommand\bibfield[2]{#2}
\providecommand\bibinfo[2]{#2}
\providecommand\natexlab[1]{#1}
\providecommand\showeprint[2][]{arXiv:#2}

\bibitem[\protect\citeauthoryear{Alam, Shaar, Dalvi, Sajjad, Nikolov, Mubarak,
  Da~San~Martino, Abdelali, Durrani, Darwish, Al-Homaid, Zaghouani, Caselli,
  Danoe, Stolk, Bruntink, and Nakov}{Alam et~al\mbox{.}}{2021}]%
        {alam-etal-2021-fighting-covid}
\bibfield{author}{\bibinfo{person}{Firoj Alam}, \bibinfo{person}{Shaden Shaar},
  \bibinfo{person}{Fahim Dalvi}, \bibinfo{person}{Hassan Sajjad},
  \bibinfo{person}{Alex Nikolov}, \bibinfo{person}{Hamdy Mubarak},
  \bibinfo{person}{Giovanni Da~San~Martino}, \bibinfo{person}{Ahmed Abdelali},
  \bibinfo{person}{Nadir Durrani}, \bibinfo{person}{Kareem Darwish},
  \bibinfo{person}{Abdulaziz Al-Homaid}, \bibinfo{person}{Wajdi Zaghouani},
  \bibinfo{person}{Tommaso Caselli}, \bibinfo{person}{Gijs Danoe},
  \bibinfo{person}{Friso Stolk}, \bibinfo{person}{Britt Bruntink}, {and}
  \bibinfo{person}{Preslav Nakov}.} \bibinfo{year}{2021}\natexlab{}.
\newblock \showarticletitle{Fighting the {COVID}-19 Infodemic: Modeling the
  Perspective of Journalists, Fact-Checkers, Social Media Platforms, Policy
  Makers, and the Society}. In \bibinfo{booktitle}{\emph{EMNLP}}.
\newblock


\bibitem[\protect\citeauthoryear{Assenmacher, Clever, Pohl, Trautmann, and
  Grimme}{Assenmacher et~al\mbox{.}}{2020}]%
        {assenmacher2020two}
\bibfield{author}{\bibinfo{person}{Dennis Assenmacher}, \bibinfo{person}{Lena
  Clever}, \bibinfo{person}{Janina~Susanne Pohl}, \bibinfo{person}{Heike
  Trautmann}, {and} \bibinfo{person}{Christian Grimme}.}
  \bibinfo{year}{2020}\natexlab{}.
\newblock \showarticletitle{A Two-Phase Framework for Detecting Manipulation
  Campaigns in Social Media}. In \bibinfo{booktitle}{\emph{SCSM}}.
\newblock


\bibitem[\protect\citeauthoryear{Barr{\'o}n-Cedeno, Jaradat, Da~San~Martino,
  and Nakov}{Barr{\'o}n-Cedeno et~al\mbox{.}}{2019}]%
        {BARRONCEDENO20191849}
\bibfield{author}{\bibinfo{person}{Alberto Barr{\'o}n-Cedeno},
  \bibinfo{person}{Israa Jaradat}, \bibinfo{person}{Giovanni Da~San~Martino},
  {and} \bibinfo{person}{Preslav Nakov}.} \bibinfo{year}{2019}\natexlab{}.
\newblock \showarticletitle{Proppy: Organizing the news based on their
  propagandistic content}.
\newblock \bibinfo{journal}{\emph{Information Processing \& Management}}
  \bibinfo{volume}{56}, \bibinfo{number}{5} (\bibinfo{year}{2019}).
\newblock


\bibitem[\protect\citeauthoryear{Bolsover and Howard}{Bolsover and
  Howard}{2017}]%
        {bolsover2017computational}
\bibfield{author}{\bibinfo{person}{G Bolsover} {and} \bibinfo{person}{P
  Howard}.} \bibinfo{year}{2017}\natexlab{}.
\newblock \showarticletitle{Computational Propaganda and Political Big Data:
  Moving Toward a More Critical Research Agenda.}
\newblock \bibinfo{journal}{\emph{Big Data}} \bibinfo{volume}{5},
  \bibinfo{number}{4} (\bibinfo{year}{2017}).
\newblock


\bibitem[\protect\citeauthoryear{Card, Boydstun, Gross, Resnik, and Smith}{Card
  et~al\mbox{.}}{2015}]%
        {card-etal-2015-media}
\bibfield{author}{\bibinfo{person}{Dallas Card}, \bibinfo{person}{Amber~E.
  Boydstun}, \bibinfo{person}{Justin~H. Gross}, \bibinfo{person}{Philip
  Resnik}, {and} \bibinfo{person}{Noah~A. Smith}.}
  \bibinfo{year}{2015}\natexlab{}.
\newblock \showarticletitle{The Media Frames Corpus: Annotations of Frames
  Across Issues}. In \bibinfo{booktitle}{\emph{ACL-IJCNLP}}.
\newblock


\bibitem[\protect\citeauthoryear{{Center for an Informed Public}, {Digital
  Forensic Research Lab}, {Graphika}, and {Stanford Internet
  Observatory}}{{Center for an Informed Public} et~al\mbox{.}}{2021}]%
        {thelongfuse2021}
\bibfield{author}{\bibinfo{person}{{Center for an Informed Public}},
  \bibinfo{person}{{Digital Forensic Research Lab}},
  \bibinfo{person}{{Graphika}}, {and} \bibinfo{person}{{Stanford Internet
  Observatory}}.} \bibinfo{year}{2021}\natexlab{}.
\newblock \showarticletitle{The Long Fuse: Misinformation and the 2020
  Election}.
\newblock \bibinfo{journal}{\emph{Stanford Digital Repository: Election
  Integrity Partnership}} (\bibinfo{year}{2021}).
\newblock


\bibitem[\protect\citeauthoryear{Conover, Ratkiewicz, Francisco,
  Gon{\c{c}}alves, Menczer, and Flammini}{Conover et~al\mbox{.}}{2011}]%
        {conover2011political}
\bibfield{author}{\bibinfo{person}{Michael Conover}, \bibinfo{person}{Jacob
  Ratkiewicz}, \bibinfo{person}{Matthew Francisco}, \bibinfo{person}{Bruno
  Gon{\c{c}}alves}, \bibinfo{person}{Filippo Menczer}, {and}
  \bibinfo{person}{Alessandro Flammini}.} \bibinfo{year}{2011}\natexlab{}.
\newblock \showarticletitle{Political polarization on {Twitter}}. In
  \bibinfo{booktitle}{\emph{AAAI ICWSM}}.
\newblock


\bibitem[\protect\citeauthoryear{Coscia and Rossi}{Coscia and Rossi}{2019}]%
        {coscia2019impact}
\bibfield{author}{\bibinfo{person}{Michele Coscia} {and} \bibinfo{person}{Luca
  Rossi}.} \bibinfo{year}{2019}\natexlab{}.
\newblock \showarticletitle{The impact of projection and backboning on network
  topologies}. In \bibinfo{booktitle}{\emph{IEEE/ACM ASONAM}}.
\newblock


\bibitem[\protect\citeauthoryear{Cresci}{Cresci}{2020}]%
        {cresci2020decade}
\bibfield{author}{\bibinfo{person}{Stefano Cresci}.}
  \bibinfo{year}{2020}\natexlab{}.
\newblock \showarticletitle{A decade of social bot detection}.
\newblock \bibinfo{journal}{\emph{Commun. ACM}} \bibinfo{volume}{63},
  \bibinfo{number}{10} (\bibinfo{year}{2020}).
\newblock


\bibitem[\protect\citeauthoryear{Da~San~Martino, Barr{\'o}n-Cedeno, Wachsmuth,
  Petrov, and Nakov}{Da~San~Martino et~al\mbox{.}}{2020a}]%
        {da2020semeval}
\bibfield{author}{\bibinfo{person}{Giovanni Da~San~Martino},
  \bibinfo{person}{Alberto Barr{\'o}n-Cedeno}, \bibinfo{person}{Henning
  Wachsmuth}, \bibinfo{person}{Rostislav Petrov}, {and}
  \bibinfo{person}{Preslav Nakov}.} \bibinfo{year}{2020}\natexlab{a}.
\newblock \showarticletitle{SemEval-2020 task 11: Detection of propaganda
  techniques in news articles}. In \bibinfo{booktitle}{\emph{SemEval}}.
\newblock


\bibitem[\protect\citeauthoryear{Da~San~Martino, Cresci, Barr\'{o}n-Cede\~{n}o,
  Yu, Di~Pietro, and Nakov}{Da~San~Martino et~al\mbox{.}}{2020b}]%
        {dasanmartino2020survey}
\bibfield{author}{\bibinfo{person}{Giovanni Da~San~Martino},
  \bibinfo{person}{Stefano Cresci}, \bibinfo{person}{Alberto
  Barr\'{o}n-Cede\~{n}o}, \bibinfo{person}{Seunghak Yu},
  \bibinfo{person}{Roberto Di~Pietro}, {and} \bibinfo{person}{Preslav Nakov}.}
  \bibinfo{year}{2020}\natexlab{b}.
\newblock \showarticletitle{{A survey on computational propaganda detection}}.
  In \bibinfo{booktitle}{\emph{IJCAI}}.
\newblock


\bibitem[\protect\citeauthoryear{Da~San~Martino, Shaar, Zhang, Yu,
  Barr{\'o}n-Cedeno, and Nakov}{Da~San~Martino et~al\mbox{.}}{2020c}]%
        {da2020prta}
\bibfield{author}{\bibinfo{person}{Giovanni Da~San~Martino},
  \bibinfo{person}{Shaden Shaar}, \bibinfo{person}{Yifan Zhang},
  \bibinfo{person}{Seunghak Yu}, \bibinfo{person}{Alberto Barr{\'o}n-Cedeno},
  {and} \bibinfo{person}{Preslav Nakov}.} \bibinfo{year}{2020}\natexlab{c}.
\newblock \showarticletitle{Prta: A system to support the analysis of
  propaganda techniques in the news}. In \bibinfo{booktitle}{\emph{ACL}}.
\newblock


\bibitem[\protect\citeauthoryear{Da~San~Martino, Yu, Barr\'{o}n-Cede\~no,
  Petrov, and Nakov}{Da~San~Martino et~al\mbox{.}}{2019}]%
        {EMNLP19DaSanMartino}
\bibfield{author}{\bibinfo{person}{Giovanni Da~San~Martino},
  \bibinfo{person}{Seunghak Yu}, \bibinfo{person}{Alberto Barr\'{o}n-Cede\~no},
  \bibinfo{person}{Rostislav Petrov}, {and} \bibinfo{person}{Preslav Nakov}.}
  \bibinfo{year}{2019}\natexlab{}.
\newblock \showarticletitle{Fine-Grained Analysis of Propaganda in News
  Articles}. In \bibinfo{booktitle}{\emph{EMNLP}}.
\newblock


\bibitem[\protect\citeauthoryear{Darwish, Stefanov, Aupetit, and Nakov}{Darwish
  et~al\mbox{.}}{2020}]%
        {darwish2020unsupervised}
\bibfield{author}{\bibinfo{person}{Kareem Darwish}, \bibinfo{person}{Peter
  Stefanov}, \bibinfo{person}{Micha{\"e}l Aupetit}, {and}
  \bibinfo{person}{Preslav Nakov}.} \bibinfo{year}{2020}\natexlab{}.
\newblock \showarticletitle{Unsupervised user stance detection on {Twitter}}.
  In \bibinfo{booktitle}{\emph{AAAI ICWSM}}.
\newblock


\bibitem[\protect\citeauthoryear{Di~Pietro, Caprolu, Raponi, and
  Cresci}{Di~Pietro et~al\mbox{.}}{2021}]%
        {dipietro2021new}
\bibfield{author}{\bibinfo{person}{Roberto Di~Pietro},
  \bibinfo{person}{Maurantonio Caprolu}, \bibinfo{person}{Simone Raponi}, {and}
  \bibinfo{person}{Stefano Cresci}.} \bibinfo{year}{2021}\natexlab{}.
\newblock \bibinfo{booktitle}{\emph{{New Dimensions of Information Warfare}}}.
  \bibinfo{series}{Advances in Information Security},
  Vol.~\bibinfo{volume}{84}.
\newblock \bibinfo{publisher}{Springer}.
\newblock


\bibitem[\protect\citeauthoryear{Dimitrov, Bin~Ali, Shaar, Alam, Silvestri,
  Firooz, Nakov, and Da~San~Martino}{Dimitrov et~al\mbox{.}}{2021a}]%
        {ACL2021:propaganda:memes}
\bibfield{author}{\bibinfo{person}{Dimitar Dimitrov}, \bibinfo{person}{Bishr
  Bin~Ali}, \bibinfo{person}{Shaden Shaar}, \bibinfo{person}{Firoj Alam},
  \bibinfo{person}{Fabrizio Silvestri}, \bibinfo{person}{Hamed Firooz},
  \bibinfo{person}{Preslav Nakov}, {and} \bibinfo{person}{Giovanni
  Da~San~Martino}.} \bibinfo{year}{2021}\natexlab{a}.
\newblock \showarticletitle{Detecting Propaganda Techniques in Memes}. In
  \bibinfo{booktitle}{\emph{ACL-IJCNLP}}.
\newblock


\bibitem[\protect\citeauthoryear{Dimitrov, Bin~Ali, Shaar, Alam, Silvestri,
  Firooz, Nakov, and Da~San~Martino}{Dimitrov et~al\mbox{.}}{2021b}]%
        {SemEval2021-6-Dimitrov}
\bibfield{author}{\bibinfo{person}{Dimiter Dimitrov}, \bibinfo{person}{Bishr
  Bin~Ali}, \bibinfo{person}{Shaden Shaar}, \bibinfo{person}{Firoj Alam},
  \bibinfo{person}{Fabrizio Silvestri}, \bibinfo{person}{Hamed Firooz},
  \bibinfo{person}{Preslav Nakov}, {and} \bibinfo{person}{Giovanni
  Da~San~Martino}.} \bibinfo{year}{2021}\natexlab{b}.
\newblock \showarticletitle{{SemEval}-2021 task 6: Detection of Persuasion
  Techniques in Texts and Images}. In \bibinfo{booktitle}{\emph{SemEval}}.
\newblock


\bibitem[\protect\citeauthoryear{Fagni and Cresci}{Fagni and Cresci}{2022}]%
        {fagni2022fine}
\bibfield{author}{\bibinfo{person}{Tiziano Fagni} {and}
  \bibinfo{person}{Stefano Cresci}.} \bibinfo{year}{2022}\natexlab{}.
\newblock \showarticletitle{Fine-Grained Prediction of Political Leaning on
  Social Media with Unsupervised Deep Learning}.
\newblock \bibinfo{journal}{\emph{Journal of Artificial Intelligence Research}}
   \bibinfo{volume}{73} (\bibinfo{year}{2022}).
\newblock


\bibitem[\protect\citeauthoryear{Ferrara, Cresci, and Luceri}{Ferrara
  et~al\mbox{.}}{2020}]%
        {ferrara2020misinformation}
\bibfield{author}{\bibinfo{person}{Emilio Ferrara}, \bibinfo{person}{Stefano
  Cresci}, {and} \bibinfo{person}{Luca Luceri}.}
  \bibinfo{year}{2020}\natexlab{}.
\newblock \showarticletitle{{Misinformation, manipulation and abuse on social
  media in the era of COVID-19}}.
\newblock \bibinfo{journal}{\emph{Journal of Computational Social Science}}
  \bibinfo{volume}{3} (\bibinfo{year}{2020}).
\newblock


\bibitem[\protect\citeauthoryear{Garimella, De~Francisci~Morales, Gionis, and
  Mathioudakis}{Garimella et~al\mbox{.}}{2017}]%
        {garimella2017reducing}
\bibfield{author}{\bibinfo{person}{Kiran Garimella}, \bibinfo{person}{Gianmarco
  De~Francisci~Morales}, \bibinfo{person}{Aristides Gionis}, {and}
  \bibinfo{person}{Michael Mathioudakis}.} \bibinfo{year}{2017}\natexlab{}.
\newblock \showarticletitle{Reducing controversy by connecting opposing views}.
  In \bibinfo{booktitle}{\emph{ACM WSDM}}.
\newblock


\bibitem[\protect\citeauthoryear{Giglietto, Righetti, Rossi, and
  Marino}{Giglietto et~al\mbox{.}}{2020}]%
        {giglietto2020takes}
\bibfield{author}{\bibinfo{person}{Fabio Giglietto}, \bibinfo{person}{Nicola
  Righetti}, \bibinfo{person}{Luca Rossi}, {and} \bibinfo{person}{Giada
  Marino}.} \bibinfo{year}{2020}\natexlab{}.
\newblock \showarticletitle{It takes a village to manipulate the media:
  coordinated link sharing behavior during 2018 and 2019 Italian elections}.
\newblock \bibinfo{journal}{\emph{Information, Communication \& Society}}
  (\bibinfo{year}{2020}).
\newblock


\bibitem[\protect\citeauthoryear{Habernal, Hannemann, Pollak, Klamm, Pauli, and
  Gurevych}{Habernal et~al\mbox{.}}{2017}]%
        {Habernal.et.al.2017.EMNLP}
\bibfield{author}{\bibinfo{person}{Ivan Habernal}, \bibinfo{person}{Raffael
  Hannemann}, \bibinfo{person}{Christian Pollak}, \bibinfo{person}{Christopher
  Klamm}, \bibinfo{person}{Patrick Pauli}, {and} \bibinfo{person}{Iryna
  Gurevych}.} \bibinfo{year}{2017}\natexlab{}.
\newblock \showarticletitle{{A}rgotario: Computational Argumentation Meets
  Serious Games}. In \bibinfo{booktitle}{\emph{EMNLP}}.
\newblock


\bibitem[\protect\citeauthoryear{Habernal, Pauli, and Gurevych}{Habernal
  et~al\mbox{.}}{2018}]%
        {Habernal2018b}
\bibfield{author}{\bibinfo{person}{Ivan Habernal}, \bibinfo{person}{Patrick
  Pauli}, {and} \bibinfo{person}{Iryna Gurevych}.}
  \bibinfo{year}{2018}\natexlab{}.
\newblock \showarticletitle{Adapting Serious Game for Fallacious Argumentation
  to {G}erman: Pitfalls, Insights, and Best Practices}. In
  \bibinfo{booktitle}{\emph{LREC}}.
\newblock


\bibitem[\protect\citeauthoryear{Horne, Dron, Khedr, and Adali}{Horne
  et~al\mbox{.}}{2018}]%
        {Horne2018}
\bibfield{author}{\bibinfo{person}{Benjamin~D. Horne}, \bibinfo{person}{William
  Dron}, \bibinfo{person}{Sara Khedr}, {and} \bibinfo{person}{Sibel Adali}.}
  \bibinfo{year}{2018}\natexlab{}.
\newblock \showarticletitle{Sampling the News Producers: A Large News and
  Feature Data Set for the Study of the Complex Media Landscape}. In
  \bibinfo{booktitle}{\emph{AAAI ICWSM}}.
\newblock


\bibitem[\protect\citeauthoryear{Jackson, Thorsen, Lilleker, and
  Weidhase}{Jackson et~al\mbox{.}}{2019}]%
        {jackson2019uk}
\bibfield{author}{\bibinfo{person}{Dan Jackson}, \bibinfo{person}{Einar
  Thorsen}, \bibinfo{person}{Darren Lilleker}, {and} \bibinfo{person}{Nathalie
  Weidhase}.} \bibinfo{year}{2019}\natexlab{}.
\newblock \bibinfo{booktitle}{\emph{{UK Election Analysis 2019: Media, Voters
  and the Campaign}}}.
\newblock \bibinfo{type}{{T}echnical {R}eport}.
  \bibinfo{institution}{Bournemouth University}.
\newblock


\bibitem[\protect\citeauthoryear{Jhaver, Boylston, Yang, and Bruckman}{Jhaver
  et~al\mbox{.}}{2021}]%
        {jhaver2021evaluating}
\bibfield{author}{\bibinfo{person}{Shagun Jhaver}, \bibinfo{person}{Christian
  Boylston}, \bibinfo{person}{Diyi Yang}, {and} \bibinfo{person}{Amy
  Bruckman}.} \bibinfo{year}{2021}\natexlab{}.
\newblock \showarticletitle{Evaluating the effectiveness of deplatforming as a
  moderation strategy on {Twitter}}. In \bibinfo{booktitle}{\emph{ACM CSCW}}.
\newblock


\bibitem[\protect\citeauthoryear{Khan, Bandyopadhyay, Ganguly, Saigal,
  Erickson~III, Protopopescu, and Ostrouchov}{Khan et~al\mbox{.}}{2007}]%
        {khan2007relative}
\bibfield{author}{\bibinfo{person}{Shiraj Khan}, \bibinfo{person}{Sharba
  Bandyopadhyay}, \bibinfo{person}{Auroop~R Ganguly}, \bibinfo{person}{Sunil
  Saigal}, \bibinfo{person}{David~J Erickson~III}, \bibinfo{person}{Vladimir
  Protopopescu}, {and} \bibinfo{person}{George Ostrouchov}.}
  \bibinfo{year}{2007}\natexlab{}.
\newblock \showarticletitle{Relative performance of mutual information
  estimation methods for quantifying the dependence among short and noisy
  data}.
\newblock \bibinfo{journal}{\emph{Physical Review E}} \bibinfo{volume}{76},
  \bibinfo{number}{2} (\bibinfo{year}{2007}).
\newblock


\bibitem[\protect\citeauthoryear{Magelinski, Ng, and Carley}{Magelinski
  et~al\mbox{.}}{2021}]%
        {magelinski2021synchronized}
\bibfield{author}{\bibinfo{person}{Thomas Magelinski},
  \bibinfo{person}{Lynnette Hui~Xian Ng}, {and} \bibinfo{person}{Kathleen~M
  Carley}.} \bibinfo{year}{2021}\natexlab{}.
\newblock \showarticletitle{A Synchronized Action Framework for Responsible
  Detection of Coordination on Social Media}.
\newblock \bibinfo{journal}{\emph{arXiv:2105.07454}} (\bibinfo{year}{2021}).
\newblock


\bibitem[\protect\citeauthoryear{Mirbabaie, Br{\"u}nker, Wischnewski, and
  Meinert}{Mirbabaie et~al\mbox{.}}{2021}]%
        {mirbabaie2021development}
\bibfield{author}{\bibinfo{person}{Milad Mirbabaie}, \bibinfo{person}{Felix
  Br{\"u}nker}, \bibinfo{person}{Magdalena Wischnewski}, {and}
  \bibinfo{person}{Judith Meinert}.} \bibinfo{year}{2021}\natexlab{}.
\newblock \showarticletitle{The Development of Connective Action during Social
  Movements on Social Media}.
\newblock \bibinfo{journal}{\emph{ACM Transactions on Social Computing}}
  \bibinfo{volume}{4}, \bibinfo{number}{1} (\bibinfo{year}{2021}).
\newblock


\bibitem[\protect\citeauthoryear{Ng, Cruickshank, and Carley}{Ng
  et~al\mbox{.}}{2021}]%
        {ng2021coordinating}
\bibfield{author}{\bibinfo{person}{Lynnette Hui~Xian Ng}, \bibinfo{person}{Iain
  Cruickshank}, {and} \bibinfo{person}{Kathleen~M Carley}.}
  \bibinfo{year}{2021}\natexlab{}.
\newblock \showarticletitle{Coordinating Narratives and the {Capitol Riots} on
  {Parler}}.
\newblock \bibinfo{journal}{\emph{arXiv:2109.00945}} (\bibinfo{year}{2021}).
\newblock


\bibitem[\protect\citeauthoryear{Nizzoli, Tardelli, Avvenuti, Cresci, and
  Tesconi}{Nizzoli et~al\mbox{.}}{2021}]%
        {nizzoli2021coordinated}
\bibfield{author}{\bibinfo{person}{Leonardo Nizzoli}, \bibinfo{person}{Serena
  Tardelli}, \bibinfo{person}{Marco Avvenuti}, \bibinfo{person}{Stefano
  Cresci}, {and} \bibinfo{person}{Maurizio Tesconi}.}
  \bibinfo{year}{2021}\natexlab{}.
\newblock \showarticletitle{Coordinated behavior on social media in 2019 UK
  general election}. In \bibinfo{booktitle}{\emph{AAAI ICWSM}}.
\newblock


\bibitem[\protect\citeauthoryear{Pacheco, Flammini, and Menczer}{Pacheco
  et~al\mbox{.}}{2020}]%
        {pacheco2020unveiling}
\bibfield{author}{\bibinfo{person}{Diogo Pacheco}, \bibinfo{person}{Alessandro
  Flammini}, {and} \bibinfo{person}{Filippo Menczer}.}
  \bibinfo{year}{2020}\natexlab{}.
\newblock \showarticletitle{Unveiling Coordinated Groups Behind White Helmets
  Disinformation}. In \bibinfo{booktitle}{\emph{WWW Companion}}.
\newblock


\bibitem[\protect\citeauthoryear{Pacheco, Hui, Torres-Lugo, Truong, Flammini,
  and Menczer}{Pacheco et~al\mbox{.}}{2021}]%
        {pacheco2020uncovering}
\bibfield{author}{\bibinfo{person}{Diogo Pacheco}, \bibinfo{person}{Pik-Mai
  Hui}, \bibinfo{person}{Christopher Torres-Lugo}, \bibinfo{person}{Bao~Tran
  Truong}, \bibinfo{person}{Alessandro Flammini}, {and}
  \bibinfo{person}{Filippo Menczer}.} \bibinfo{year}{2021}\natexlab{}.
\newblock \showarticletitle{Uncovering coordinated networks on social media:
  Methods and case studies}. In \bibinfo{booktitle}{\emph{AAAI ICWSM}}.
\newblock


\bibitem[\protect\citeauthoryear{Pei, Muchnik, Andrade~Jr, Zheng, and
  Makse}{Pei et~al\mbox{.}}{2014}]%
        {pei2014searching}
\bibfield{author}{\bibinfo{person}{Sen Pei}, \bibinfo{person}{Lev Muchnik},
  \bibinfo{person}{Jos{\'e}~S Andrade~Jr}, \bibinfo{person}{Zhiming Zheng},
  {and} \bibinfo{person}{Hern{\'a}n~A Makse}.} \bibinfo{year}{2014}\natexlab{}.
\newblock \showarticletitle{Searching for superspreaders of information in
  real-world social media}.
\newblock \bibinfo{journal}{\emph{Scientific reports}}  \bibinfo{volume}{4}
  (\bibinfo{year}{2014}).
\newblock


\bibitem[\protect\citeauthoryear{Rashkin, Choi, Jang, Volkova, and
  Choi}{Rashkin et~al\mbox{.}}{2017}]%
        {rashkin-EtAl:2017:EMNLP2017}
\bibfield{author}{\bibinfo{person}{Hannah Rashkin}, \bibinfo{person}{Eunsol
  Choi}, \bibinfo{person}{Jin~Yea Jang}, \bibinfo{person}{Svitlana Volkova},
  {and} \bibinfo{person}{Yejin Choi}.} \bibinfo{year}{2017}\natexlab{}.
\newblock \showarticletitle{Truth of Varying Shades: Analyzing Language in Fake
  News and Political Fact-Checking}. In \bibinfo{booktitle}{\emph{EMNLP}}.
\newblock


\bibitem[\protect\citeauthoryear{Rauchfleisch and Kaiser}{Rauchfleisch and
  Kaiser}{2020}]%
        {rauchfleisch2020false}
\bibfield{author}{\bibinfo{person}{Adrian Rauchfleisch} {and}
  \bibinfo{person}{Jonas Kaiser}.} \bibinfo{year}{2020}\natexlab{}.
\newblock \showarticletitle{The false positive problem of automatic bot
  detection in social science research}.
\newblock \bibinfo{journal}{\emph{PLoS One}} \bibinfo{volume}{15},
  \bibinfo{number}{10} (\bibinfo{year}{2020}).
\newblock


\bibitem[\protect\citeauthoryear{Sayyadiharikandeh, Varol, Yang, Flammini, and
  Menczer}{Sayyadiharikandeh et~al\mbox{.}}{2020}]%
        {sayyadiharikandeh2020detection}
\bibfield{author}{\bibinfo{person}{Mohsen Sayyadiharikandeh},
  \bibinfo{person}{Onur Varol}, \bibinfo{person}{Kai-Cheng Yang},
  \bibinfo{person}{Alessandro Flammini}, {and} \bibinfo{person}{Filippo
  Menczer}.} \bibinfo{year}{2020}\natexlab{}.
\newblock \showarticletitle{Detection of novel social bots by ensembles of
  specialized classifiers}. In \bibinfo{booktitle}{\emph{ACM CIKM}}.
\newblock


\bibitem[\protect\citeauthoryear{Schumacher}{Schumacher}{2019}]%
        {schumacher2019brexit}
\bibfield{author}{\bibinfo{person}{Shannon Schumacher}.}
  \bibinfo{year}{2019}\natexlab{}.
\newblock \bibinfo{booktitle}{\emph{{Brexit divides the UK, but partisanship
  and ideology are still key factors}}}.
\newblock \bibinfo{type}{{T}echnical {R}eport}. \bibinfo{institution}{Pew
  Research Center}.
\newblock


\bibitem[\protect\citeauthoryear{Serrano, Bogun{\'a}, and Vespignani}{Serrano
  et~al\mbox{.}}{2009}]%
        {serrano2009extracting}
\bibfield{author}{\bibinfo{person}{M~{\'A}ngeles Serrano},
  \bibinfo{person}{Mari{\'a}n Bogun{\'a}}, {and} \bibinfo{person}{Alessandro
  Vespignani}.} \bibinfo{year}{2009}\natexlab{}.
\newblock \showarticletitle{{Extracting the multiscale backbone of complex
  weighted networks}}.
\newblock \bibinfo{journal}{\emph{PNAS}} \bibinfo{volume}{106},
  \bibinfo{number}{16} (\bibinfo{year}{2009}).
\newblock


\bibitem[\protect\citeauthoryear{Sharma, Zhang, Ferrara, and Liu}{Sharma
  et~al\mbox{.}}{2021}]%
        {sharma2021identifying}
\bibfield{author}{\bibinfo{person}{Karishma Sharma}, \bibinfo{person}{Yizhou
  Zhang}, \bibinfo{person}{Emilio Ferrara}, {and} \bibinfo{person}{Yan Liu}.}
  \bibinfo{year}{2021}\natexlab{}.
\newblock \showarticletitle{Identifying coordinated accounts on social media
  through hidden influence and group behaviours}. In
  \bibinfo{booktitle}{\emph{ACM KDD}}.
\newblock


\bibitem[\protect\citeauthoryear{Starbird, Arif, and Wilson}{Starbird
  et~al\mbox{.}}{2019}]%
        {starbird2019disinfo}
\bibfield{author}{\bibinfo{person}{Kate Starbird}, \bibinfo{person}{Ahmer
  Arif}, {and} \bibinfo{person}{Tom Wilson}.} \bibinfo{year}{2019}\natexlab{}.
\newblock \showarticletitle{{Disinformation as Collaborative Work: Surfacing
  the Participatory Nature of Strategic Information Operations}}. In
  \bibinfo{booktitle}{\emph{ACM CSCW}}.
\newblock


\bibitem[\protect\citeauthoryear{Tardelli, Avvenuti, Tesconi, and
  Cresci}{Tardelli et~al\mbox{.}}{2022}]%
        {tardelli2022detecting}
\bibfield{author}{\bibinfo{person}{Serena Tardelli}, \bibinfo{person}{Marco
  Avvenuti}, \bibinfo{person}{Maurizio Tesconi}, {and} \bibinfo{person}{Stefano
  Cresci}.} \bibinfo{year}{2022}\natexlab{}.
\newblock \showarticletitle{Detecting inorganic financial campaigns on
  Twitter}.
\newblock \bibinfo{journal}{\emph{Information Systems}} (\bibinfo{year}{2022}).
\newblock


\bibitem[\protect\citeauthoryear{Trujillo and Cresci}{Trujillo and
  Cresci}{2022}]%
        {trujillo2022make}
\bibfield{author}{\bibinfo{person}{Amaury Trujillo} {and}
  \bibinfo{person}{Stefano Cresci}.} \bibinfo{year}{2022}\natexlab{}.
\newblock \showarticletitle{Make Reddit Great Again: Assessing Community
  Effects of Moderation Interventions on r/The\_Donald}.
\newblock \bibinfo{journal}{\emph{arXiv:2201.06455}} (\bibinfo{year}{2022}).
\newblock


\bibitem[\protect\citeauthoryear{Vargas, Emami, and Traynor}{Vargas
  et~al\mbox{.}}{2020}]%
        {vargas2020detection}
\bibfield{author}{\bibinfo{person}{Luis Vargas}, \bibinfo{person}{Patrick
  Emami}, {and} \bibinfo{person}{Patrick Traynor}.}
  \bibinfo{year}{2020}\natexlab{}.
\newblock \showarticletitle{On the detection of disinformation campaign
  activity with network analysis}. In \bibinfo{booktitle}{\emph{ACM CCSW}}.
\newblock


\bibitem[\protect\citeauthoryear{Wardle and Derakhshan}{Wardle and
  Derakhshan}{2017}]%
        {wardle2017information}
\bibfield{author}{\bibinfo{person}{Claire Wardle} {and}
  \bibinfo{person}{Hossein Derakhshan}.} \bibinfo{year}{2017}\natexlab{}.
\newblock \showarticletitle{Information disorder: Toward an interdisciplinary
  framework for research and policy making}.
\newblock \bibinfo{journal}{\emph{Council of Europe}}  \bibinfo{volume}{27}
  (\bibinfo{year}{2017}).
\newblock


\bibitem[\protect\citeauthoryear{Weber and Neumann}{Weber and Neumann}{2020}]%
        {weber2020s}
\bibfield{author}{\bibinfo{person}{Derek Weber} {and} \bibinfo{person}{Frank
  Neumann}.} \bibinfo{year}{2020}\natexlab{}.
\newblock \showarticletitle{Who's in the Gang? Revealing Coordinating
  Communities in Social Media}. In \bibinfo{booktitle}{\emph{IEEE/ACM ASONAM}}.
\newblock


\bibitem[\protect\citeauthoryear{Weber and Neumann}{Weber and Neumann}{2021}]%
        {weber2021amplifying}
\bibfield{author}{\bibinfo{person}{Derek Weber} {and} \bibinfo{person}{Frank
  Neumann}.} \bibinfo{year}{2021}\natexlab{}.
\newblock \showarticletitle{Amplifying influence through coordinated behaviour
  in social networks}.
\newblock \bibinfo{journal}{\emph{SNAM}} (\bibinfo{year}{2021}).
\newblock


\bibitem[\protect\citeauthoryear{Zannettou, Caulfield, Blackburn,
  De~Cristofaro, Sirivianos, Stringhini, and Suarez-Tangil}{Zannettou
  et~al\mbox{.}}{2018}]%
        {zannettou2018origins}
\bibfield{author}{\bibinfo{person}{Savvas Zannettou}, \bibinfo{person}{Tristan
  Caulfield}, \bibinfo{person}{Jeremy Blackburn}, \bibinfo{person}{Emiliano
  De~Cristofaro}, \bibinfo{person}{Michael Sirivianos},
  \bibinfo{person}{Gianluca Stringhini}, {and} \bibinfo{person}{Guillermo
  Suarez-Tangil}.} \bibinfo{year}{2018}\natexlab{}.
\newblock \showarticletitle{On the origins of memes by means of fringe {Web}
  communities}. In \bibinfo{booktitle}{\emph{ACM IMC}}.
\newblock


\bibitem[\protect\citeauthoryear{Zannettou, Caulfield, De~Cristofaro,
  Kourtellis, Leontiadis, Sirivianos, Stringhini, and Blackburn}{Zannettou
  et~al\mbox{.}}{2017}]%
        {zannettou2017web}
\bibfield{author}{\bibinfo{person}{Savvas Zannettou}, \bibinfo{person}{Tristan
  Caulfield}, \bibinfo{person}{Emiliano De~Cristofaro},
  \bibinfo{person}{Nicolas Kourtellis}, \bibinfo{person}{Ilias Leontiadis},
  \bibinfo{person}{Michael Sirivianos}, \bibinfo{person}{Gianluca Stringhini},
  {and} \bibinfo{person}{Jeremy Blackburn}.} \bibinfo{year}{2017}\natexlab{}.
\newblock \showarticletitle{The {Web} centipede: understanding how web
  communities influence each other through the lens of mainstream and
  alternative news sources}. In \bibinfo{booktitle}{\emph{ACM IMC}}.
\newblock


\bibitem[\protect\citeauthoryear{Zhang, Da~San~Martino, Barr{\'o}n-Cedeno,
  Romeo, An, Kwak, Staykovski, Jaradat, Karadzhov, Baly, Darwish, Glass, and
  Nakov}{Zhang et~al\mbox{.}}{2019}]%
        {zhang2019tanbih}
\bibfield{author}{\bibinfo{person}{Yifan Zhang}, \bibinfo{person}{Giovanni
  Da~San~Martino}, \bibinfo{person}{Alberto Barr{\'o}n-Cedeno},
  \bibinfo{person}{Salvatore Romeo}, \bibinfo{person}{Jisun An},
  \bibinfo{person}{Haewoon Kwak}, \bibinfo{person}{Todor Staykovski},
  \bibinfo{person}{Israa Jaradat}, \bibinfo{person}{Georgi Karadzhov},
  \bibinfo{person}{Ramy Baly}, \bibinfo{person}{Kareem Darwish},
  \bibinfo{person}{James Glass}, {and} \bibinfo{person}{Preslav Nakov}.}
  \bibinfo{year}{2019}\natexlab{}.
\newblock \showarticletitle{Tanbih: Get To Know What You Are Reading}. In
  \bibinfo{booktitle}{\emph{EMNLP}}.
\newblock


\end{thebibliography}

\end{document}